\documentclass[preprints,article,accept,pdftex,moreauthors]{Definitions/mdpi} 

\usepackage{amsmath}
\usepackage{placeins}
\usepackage{listings} 
\firstpage{1} 
\pubvolume{1}
\issuenum{1}
\articlenumber{0}
\pubyear{2024}
\copyrightyear{2024}
\datereceived{ } 
\daterevised{ } % Comment out if no revised date
\dateaccepted{ } 
\datepublished{ } 
\hreflink{https://doi.org/} % If needed use \linebreak
\pdfoutput=1 % Uncommented for upload to arXiv.org
\Title{Cross$^\dagger$RT: A cross platform programming technology for hardware-accelerated ray tracing in CG and CV applications}

\TitleCitation{Title}

\Author{Vladimir Frolov $^{1\ddagger}$*\orcidA{}, Vadim Sanzharov $^{1,\ddagger}$\orcidB{}, Garifullin Albert $^{2}$ \orcidC{}, Maxim Raenchuk $^{1}$ \orcidD and Alexei Voloboy $^{2}$\orcidD}

\AuthorNames{Vladimir Frolov, Vadim Sanzharov, Garifullin Albert, Maxim Raenchuk and Alexei Voloboy}

\AuthorCitation{Frolov et al.}
\address{%
$^{1}$ \quad IAI Lomonosov Moscow State University, GSP-1, Lomonosovsky Prospect, 1, 27, Moscow, 119191, Russia\\
$^{2}$ \quad Keldysh Institute of Applied Mathematics, Miusskaya sq., 4, Moscow, 125047, Russia}

\corres{Correspondence: v.frolov@iai.msu.ru; vladimir.frolov@graphics.cs.msu.ru. Tel.: +7(916)8114565}

\secondnote{These authors contributed equally to this work.}
\abstract{We propose a programming technology that bridges the gap between cross-platform compatibility and hardware acceleration in ray tracing applications. We achieve this through a proposed methodology, where the developer defines the algorithm, and the translator manages its implementation details on specific hardware or APIs. There are two distinctive features: on one hand, our translator is capable of generating hardware-accelerated code from 100\% hardware-agnostic, object-oriented algorithm description in C++; on the other hand, it allows the user to define a software fallback for hardware functionality, enabling the code to run on CPUs and GPUs that lack hardware acceleration. The output is an implementation of the user's algorithm via GPU programming API, closely resembling a manually ported version from pure C++. The generated code is highly editable and readable, allowing users to enhance it with additional hardware acceleration not initially included. This flexibility ensures access to virtually any hardware capability, beyond those directly supported by our generator. Our translator can generate path tracing implementations as a single megakernel (which is often used in computer graphics) or as multiple kernels (called plain wavefront, which is often used in computer vision for inverse rendering), without altering the programming model and input source code for the user. The wavefront mode is essential for neural radiance fields (NeRF) and neural Signed Distance Functions (SDF), as their efficient evaluation is not feasible with independent threads. We have validated our technology on several tasks related to ray tracing: (1) parallel BVH tree build and traversal (for triangle meshes), (2) ray-surface intersection for SDF, (3) ray-volume intersection for ReLU-Fields, and (4) Path Tracing with complex material models. Each of these tasks was implemented as separate lightweight C++ library without any hardware specifics and then processed with our technology. Our experiments have demonstrated that for tasks (1--3), we achieve performance levels on GPUs comparable to optimized manual implementations written by experts in the field for both HW-accelerated and software-based implementations. For Path Tracing we outperform existing implementations. }


\begin{document}

%%%%%%%%%%%%%%%%%%%%%%%%%%%%%%%%%%%%%%%%%%
\setcounter{section}{0} %% Remove this when starting to work on the template.
\section{Introduction}

Existing ray tracing technologies can be divided into two categories: (1) libraries and frameworks that implement ray-geometry or ray-volume intersection algorithms (which is typically closely related to algorithms for parallel construction of acceleration structures, such as BVH trees \cite{HPLOC}), and (2) programming technologies designed both for implementing intersection algorithms and for building complete solutions on top of it. For example, Nvidia OptiX \cite{optix} is a programming technology which allows porting the whole algorithms written in C++ to GPU while OptiX Prime \cite{optixprime} is a ray tracing library for CUDA which involves separate implementation of a user algorithm on CUDA. 

Libraries may include multiple implementations for different hardware platforms, achieving cross-platform compatibility. However, such solutions are typically inflexible since different backends are often developed independently. It is typical for CPU backends to have more features compared to GPU ones (for example, OpenVDB \cite{OpenVDB} and its GPU version NanoVDB), as using GPUs involves several challenges: 

\begin{enumerate}
\item The developer must explicitly use a specific GPU programming API or framework, requiring the entire program to be written within that framework (which in fact results in a loss of cross-platform compatibility).  A CPU <=> GPU data transfer negates the entire performance gain. In fact such a transfer is expensive enough even from GPU multiprocessor to DRAM and vice versa.
\item The developer must explicitly manage data transfers between library and user code in separate buffers and implement program logic within the separate compute kernel execution paradigm. This is not only inconvenient but also imposes problems, such as the inability to implement recursion and increased memory consumption.
\end{enumerate}

Cross-platform support in such solutions can be achieved through API wrappers, such as \cite{CSharpBackEnd} or \cite{sokol}. However, this forces users to develop applications in a heterogeneous environment, where computational logic is described in a shader language, while the rest of the program interacts with the underlying API through a wrapper. This approach is notoriously difficult to debug.

In contrast to ray tracing libraries, programming technologies provide a more general and accessible solution. It simplifies the integration of ray tracing into applications by handling much of the complexity internally, allowing developers to focus on higher-level algorithm design. Due to both flexibility and performance issues, programming technologies currently dominate and are widely used for different applications such as optical simulators \cite{opticks}, radar design \cite{OptiXRadar}, physically based rendering \cite{pbrt4,Luisa}, visualization \cite{OspRaySitu}, 3D reconstruction via inverse rendering \cite{mitsuba2,diffSDF2022}, neutron transport \cite{neutrons} and many others. 

Unfortunately, most of popular ray tracing programming technologies are tightly coupled to the hardware of a specific vendor due to involvement of compilation process. For example, OptiX \cite{optix} targets Nvidia hardware while Embree\cite{Embree} + SYCL \cite{sycl} bundle depends on Intel. Existing exceptions (such as WebRays \cite{WebRays}), unfortunately, and as expected, lack the necessary hardware support required today. We will cover this problem in detail within the related  section.

\subsection{Virtual functions}

One of the most challenging problems in ray tracing applications is the diversity of light-matter interaction models. A single scene may utilize both relatively simple and well-known microfacet models (for example, \cite{GGX}) and significantly more complex, computationally intensive ones like multi-layered material with internal scattering in PBRT4 \cite{pbrt4}. Traditionally, on CPUs, the flexibility required to support such a variety of models is typically achieved through the use of virtual functions. However, while the implementation of virtual functions on CPUs generally does not present significant difficulties, the same cannot be said for GPUs. On a GPU, virtual functions are not merely a tool for achieving flexibility, they are a mechanism for efficiently executing complex code with heterogeneous computational workloads. Moreover, they represent one of the most challenging hardware-accelerated features to implement, and are available on only a limited number of GPUs.

Our solution addresses this issue with particular attention. In our translator, several options are provided that allow the generation of code (1) with virtual function calls in the ray tracing pipeline and (2) with software emulation of virtual function calls in compute shaders. The latter enables optimization of code for GPUs without support for ray tracing pipeline (current mobile GPUs for example) based on specialization constants in Vulkan.

%%%%%%%%%%%%%%%%%%%%%%%%%%%%%%%%%%%%%%%%%%
\section{Related work}

\subsection{Parallel execution models}

When implementing massively parallel computations on a GPU (same for vectorized code on CPU), there are three fundamental models: megakernel, plain wavefront (also sometimes called separate kernels), and wavefront \cite{MegakernelHarmful}.

\begin{enumerate}
\item In the \textbf{megakernel} model, all computations are implemented in a single large compute shader. This model is the most flexible and currently the most performant for code with relatively low register pressure. 
\item In the \textbf{plain wavefront} model (also called \textbf{separate kernel}), computations are divided into an independent set of compute kernels, which are generally executed sequentially, with intermediate data stored in DRAM.  
\item The \textbf{wavefront} model, on the other hand, implements computations using a set of queues organized according to the type of program being executed. The GPU fetches work from these queues in blocks, executes it, and fills other queues with the results of the current program's execution. The actual implementation of this model may vary. For example, prior to the introduction of hardware-accelerated ray tracing pipelines (RT pipeline), such a model was implemented through a state machine in a megakernel, where each state was defined by a so-called RT program \cite{MegakernelHarmful,optix}. Wavefront model is used in PBRT4 \cite{pbrt4}.
\end{enumerate}

\subsection{RT pipeline}

Technologies from this group implement a programming model with two key functionalities: virtual function calls (for handling rays intersecting surfaces with different materials or geometries) and recursive virtual function calls, enabling further ray tracing. While this model suits recursive Whitted-style ray tracing \cite{WhittedRT}, used for refractive objects, practical applications rarely need it. Computational applications typically use single reflected rays and stochastic algorithms, while gaming applications limit recursion depth or avoid it. Additionally, many GPUs don't support recursion depths beyond 1. Even when recursion is necessary, it’s often cheaper to use software implementation via stack, where only required state elements are saved, such as in ray tracing inside diamonds, where recursion depth is high but the surface never changes material properties. 

\textbf{OptiX} \cite{optix} is the most famous ray tracing programming technology for GPU. Currently, it is one of the few programming technologies that allows writing algorithms in C++ which can than run on both CPU and GPU (although, some parts will still be GPU or CPU specific). For example, PBRT4 \cite{pbrt4} actively leverages this capability. Therefore, despite being Nvidia's proprietary technology, OptiX is somewhat cross-platform. However, there are several nuances to working with OptiX. An efficient implementation of a ray tracing application using OptiX requires the use of a specialized programming model (RT pipeline), which does not directly map to CPU code. Working with RT pipeline is non-trivial, therefore some helper libraries such as OWL exist \cite{OWL}. For instance, PBRT4 does not employ this model but instead implements all computations using manually implemented wavefront approach. Additionally, OptiX is only compatible with Nvidia GPUs and does not have a CPU fallback. Developer needs to implement a CPU renderer separately. 

\textbf{Vulkan} \cite{vulkan} and DirectX12 \cite{DXR} are cross-platform low-level programming technologies that support ray tracing pipeline. However, development in Vulkan is approximately five times more labor-intensive \cite{VulkanIsComplex} compared to OptiX, making it accessible to only a limited number of developers. For example, Nvidia's in-house Falcor \cite{falcor} framework is implemented with two backends (Vulkan and DirectX12) and employs a complex code generation mechanism, a custom API built on top of Vulkan and DirectX12, as well as their own shader language called SLANG \cite{SLANGD}. One of the key challenges faced by APIs such as Vulkan is their attempt to combine two inherently conflicting goals: cross-platform compatibility and hardware acceleration. This results in an expected overcomplication of the API, manifesting in cumbersome low-level programming through abstract tools. Consequently, even the simplest examples of ray tracing or graphics pipelines in Vulkan can require several thousand lines of source code, whereas other APIs may achieve similar functionality with only a few lines of code (two order of magnitude less in fact). It also should be noted, that some ray tracing features (such as motion blur) are available in Vulkan only as vendor-specific extensions.

\subsection{JIT compilers}

\textbf{Dr.JIT} \cite{drjit} is one of the world's first JIT compilers for GPUs, distinguished by its ability to perform automatic differentiation, which would be nearly impossible without dynamic execution flow analysis. DrJIT processes code written in C++ or Python and generates implementations for vectorized CPU instructions via LLVM, or for GPUs via CUDA and OptiX. It supports hardware-accelerated ray-surface intersection queries exclusively with triangular primitives, as it does not employ a ray-tracing pipeline. Notably, DrJIT offers a unique solution to the challenge of virtual function calls. Instead of relying on the ray-tracing pipeline, DrJIT performs full devirtualization of functions based on current control flow analysis, enabling the execution of complex code in a manner distinct from other technologies.

However, a significant limitation of DrJIT is its reliance on a tensor programming model, similar to PyTorch \cite{pytorch}. This model inherently requires programming at the level of large data arrays in a plain wavefront style, which introduces constraints and can be inconvenient. For instance, instead of using the type float, the programmer must work with CUDAArray<float>. This is one reason why DrJIT programs, like those in PyTorch, tend to be memory-intensive. Nonetheless, one of DrJIT's key features is its automatic conversion of plain wavefronts into megakernels and optimizations achievable only at runtime, which is the primary reason for the high performance of the Mitsuba3 \cite{mitsuba3} rendering system, for which DrJIT was developed. While DrJIT represents one of the most advanced approaches, it is also memory-intensive, complex, and costly to implement and debug.

\textbf{Luisa} \cite{Luisa} unlike DrJIT is not focused on automatic differentiation and is much more cross-platform, supporting a wide range of backends, including CUDA, DirectX, Metal and ISPC. Luisa does not employ a tensor programming model; instead, user code is written in standard C++. The system operates by translating C++ code at the level of individual AST (Abstract Syntax Tree) elements, an unusual design choice (different to analogues: Halide \cite{halide} an Taichi \cite{taichi}). Additionally, an interesting feature of Luisa is its implementation of polymorphism and virtual functions. Luisa implements two strategies for polymorphism: (1) de-virtualized host-side polymorphism and (2) device-side dynamic dispatch. The first strategy is a form of static polymorphism based on function and data type parameterization of individual computational kernels, analogous to advanced C++ templates for kernels. The second strategy involves generating switch constructs within shaders to implement dynamic polymorphism. While static polymorphism is a convenient tool, it primarily serves as a matter of convenience. Dynamic polymorphism, on the other hand, represents concrete functionality, supported in various ways by hardware. Therefore, we place greater emphasis on this aspect.

In our view, Luisa implements a semi-JIT approach rather than a true JIT compiler. The difference lies in that a full JIT compiler performs control flow analysis on the GPU and generates computational kernels based on the actual computations being executed. Luisa, however, defines templated kernels, which are generated/specialized on the CPU side before the actual GPU execution begins. This approach allows, for example, the removal of unused classes from switch statements when such information is known in advance from the scene which is explicitly mentioned in \cite{Luisa}: ``Thus, only the necessary kernels need to be dynamically generated and compiled, rather than the numerous combinations of all shader variants ... it allows generating code only for used
subtypes, which potentially reduces code size and register usage, providing better performance.''  

%Finally, it is important to note that this approach is not very convenient for ISPC, which, unlike DirectX12 and Vulkan, requires Ahead Of Time (AoT) compilation for kernels on the developer's machine. Therefore, using the Luisa-JIT approach with ISPC means bundling ISPC with Luisa and compiling the entire program via LLVM/MCJIT \cite{MCJIT} at runtime. Unfortunately, on some platforms (e.g., Android), this is nearly impossible due to security requirements.

\subsection{Other related to ray tracing}

\textbf{WebRays} \cite{WebRays} is a cross-platform, GPU-accelerated ray intersection framework for the World Wide Web. It is designed as a "thin-as-possible abstraction" layer built on top of WebGL 2.0. Device-side computation in WebGL 2.0 is restricted to plain fragment shaders, which introduces limitations, particularly the absence of direct hardware acceleration. WebRays supports both megakernel and plain wavefront implementations. However, these implementations are not available in an automatic or semi-automatic fashion. Users are required to manually implement a megakernel using GLSL (invoking ray query functions) or to implement the plain wavefront approach through the JavaScript API, executing the ray intersection procedure as a separate rectangle draw pass. Although the programming model suggests that hardware support for ray-surface intersection might become available in a future version, potentially implemented on top of another API (e.g., DirectX12 or Vulkan), it remains significantly more constrained compared to the ray tracing pipeline-based technologies discussed earlier. The WebRays programming model does not support arbitrary geometric types, virtual functions, or recursion. This exemplifies the typical limitations faced by existing cross-platform GPU programming technologies, such as OpenCL.

\textbf{ISPC} \cite{ISPC} is acronym for Intel Single Program Multiple Data Compiler. Originally, it was designed for CPU code vectorization but later Intel GPU support was also added. A notable strength of ISPC is its ability to generate code utilizing CPU vector instructions and directly access host memory pointers. This can accelerate algorithms when data transfer to and from the GPU is prohibitively expensive. However, a key limitation of ISPC is its support for only C99 in kernels, which, similar to OpenCL, often requires substantial code modifications. As a result, in \cite{IPSCPath}, only several parts of the rendering pipeline were rewritten using ISPC.

\textbf{SYCL} \cite{sycl} is a programming model providing an abstraction layer which allows writing C++ code for execution on different hardware  platforms (such as CPUs, GPUs and FPGAs). One of the most popular implementations of SYCL is DPC++ ("data parallel C++") compiler initially developed by Intel. Extensions for DPC++ were developed introducing support for Nvidia \cite{oneapi-nv} and AMD GPUs \cite{oneapi-nv}, Huawei Ascend AI chips \cite{dpc-huawei}. There also other SYCL implementations supporting different platforms \cite{sycl-adaptive-cpp} and research trying to introduce Vulkan backend for SYCL \cite{Sylkan}. 

At the moment, SYCL is a technology for general purpose computations and thus generally can't make use of highly specific hardware features such as ray-tracing. However, well known ray tracing library \textbf{Embree} \cite{embree}, which initially targeted CPUs by making heavy use of vectorization (including support for ISPC \cite{ispc}), starting with version 4.0 added support for SYCL technology allowing to use it on Intel ARC GPUs. However, only Intel oneAPI DPC++ compiler and Intel ARC GPUs are supported. 
Embree focuses only on acceleration structure building and ray traversal functionalities with support for variety of features and geometry types (such as curves, subdivision surfaces, custom geometry etc.). Being a standalone library, Embree can be relatively easily integrated with user applications and is widely used in rendering systems for CPU backends, but GPU support via SYCL is vendor specific at the moment. 

There is some potential for SYCL to support cross-platform ray tracing. For example, \cite{sycl-adaptive-cpp} can output SPIR-V representation, which it uses for Intel GPU targets, and SPIR-V can potentially include hardware ray tracing instructions. It is likely that DPC++ compiler does this under the hood for Intel GPUs. However, for cross-platform compatibility some sort of standardized ray tracing interface in SYCL C++ would be required.

\textbf{HIPRT} \cite{hiprt} is a ray-tracing framework implemented using HIP technology \cite{hip}, and thus can run on AMD and Nvidia GPUs. However, ray traversal is hardware-accelerated only on new AMD GPUs (RDNA2 and RDNA3). Similar to Embree, HIPRT focuses on acceleration structure building and ray traversal algorithms, providing several build algorithms (with different construction and trace performance trade-offs) and features such as multi-level instancing, custom primitives and variety of motion blur techniques. As opposed to OptiX, HIPRT doesn't have an equivalent to ray tracing pipeline or any shading functionality. This makes it a relatively lightweight solution which is easier to use and integrate with user applications. Another interesting feature is the ability to import custom prebuilt BVH trees into the framework, which allows user to use a custom build algorithm with HIPRT.
It should be noted, that support for custom primitives for ray tracing is also a case for virtual functions. HIPRT implements custom primitives and intersection filters by generating dispatching function from user-supplied intersection functions for different ray types at compile time to leverage optimization capabilities of the compiler.
Overall, HIPRT is flexible, relatively small framework that has somewhat better cross-platform compatibility than other technologies, although it is still limited to desktop (and to some extent server) GPUs while hardware-acceleration is supported only for specific AMD GPUs. 

\textbf{kernel\_slicer} \cite{kslicer} is source-to-source translator which takes C++ class as input and generates implementation of an algorithms from this class in Vulkan and ISPC as output. It has ability to generate ray query calls in compute shader. This approach significantly differs from previous ones by generating readable representations of both host (C++) and device code (C99 for ISPC, GLSL for Vulkan). This allows the generated code to be integrated into the traditional development pipeline, as it is functionally similar to manually written code: developers can modify, refine, debug, and replace individual computational kernels and other functions using traditional inheritance and virtual function mechanisms in the generated host code. In other words, when working with kernel\_slicer, polymorphism is mainly applied at the host code level. An early version of kernel\_slicer had some experiments with dynamic polymorphism  \cite{kslicer2021}, however, the implementation was done via clspv complier \cite{clspv} (OpenCL shaders for Vulkan) as pointers and the pointer cast is available in clspv memory model, but not available in GLSL. At the same time, this is a significant limitation for hardware-accelerated features such as ray queries which is not supported (and not planned) in clspv.

Another significant distinction of kernel\_slicer is the absence of a dedicated API for host and device code interaction. It generates a class with the same API as the base class. Consequently, the user defines this API during the development of the base class code, and the generated code is integrated by simply swapping the implementation pointer from original class to the generated one.

\subsection{Other essential technologies}

There are several approaches that are not directly related to ray tracing but utilize similar concepts. The main issue with these approaches is the lack of ray tracing hardware acceleration support and the absence of mechanisms to implement such support in the future. We will discuss them further.

\textbf{Circle} \cite{circle} is a C++ compiler which allows C++ to SPIR-V shader compilation with support for graphics and ray tracing pipelines. Unfortunately, this approach merely replaces GLSL shaders with C++ shaders without creating a truly homogeneous system. The user is still required to write all host Vulkan code for resource management, pipeline creation, and other tasks manually. 

Approach introduced in \cite{UE4Cpp} allows the creation of shaders for the graphics pipeline in C++ within a unified framework where host and device code is mixed. A distinctive feature of \cite{UE4Cpp} is the use of C++ class inheritance to integrate generated code with existing code and for static dispatching of shaders and shader specialization. This makes the ideas in \cite{UE4Cpp} strongly related to kernel\_slicer \cite{kslicer} and ours. Another feature common to our work is that \cite{UE4Cpp} allows the use of HLSL on top of the generated code, enabling direct usage in cases where some hardware features are not supported by their translator. 

\subsection{Related work summary}

Ray-tracing technologies, which are in any way affiliated with hardware vendors, provide hardware acceleration only for the affiliated platform: HIPRT for modern AMD RDNA2 and RDNA3 GPUs, Embree version 4.x+ for Intel ARC GPUs, OptiX for Nvidia GPUs. Which is understandable because no one wants to do the work for the benefit of their competitors. For independent solutions this leads to necessity to implement several backends for different hardware platforms which is very expensive, time-consuming and often leads to cumbersome and hard-to-use results. Vulkan could be a "one backend to rule them all", but its inherent complexity and limitations of shading languages (in particular, GLSL) turn developers away from it. And thus there is no truly cross-platform ready-to-use solution for professional ray-tracing applications.

We propose \textbf{Cross$^\dagger$RT}: a solution that addresses this gap, enabling developers to remain independent of specific hardware platforms while maximizing the utilization of hardware acceleration on those platforms where it is available. Our work is based on kernel\_slicer and represents its further development. Among other technologies, the closest counterpart is Luisa. In section \ref{contribution} we will detail the differences between these technologies and our proposed approach.

\section{Proposed solution}

\begin{figure}[h]
	\includegraphics[width=\linewidth]{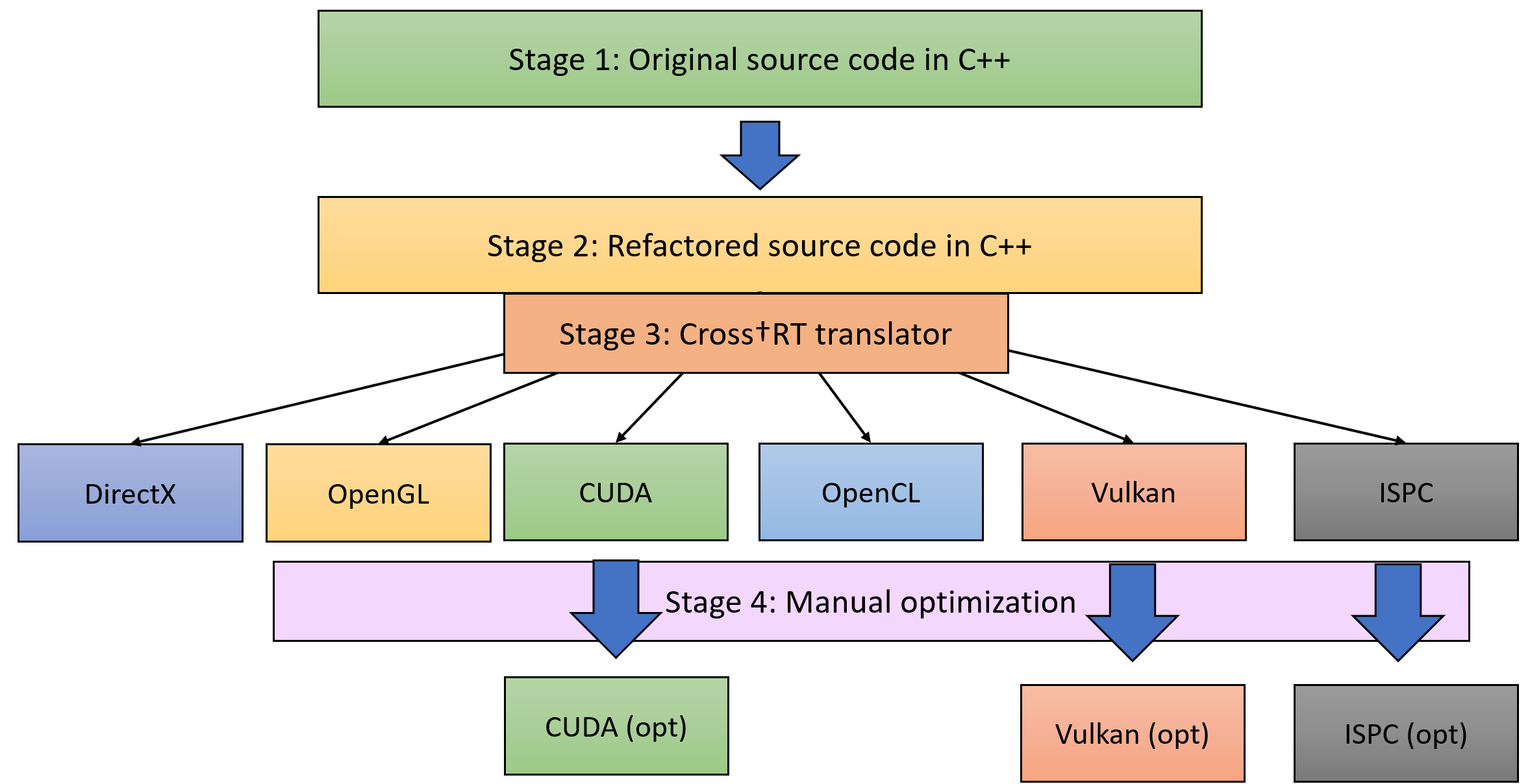}
	\caption{The workflow with the proposed technology follows these steps. In the first stage, the programmer develops their algorithm in C++ without any restrictions. In the second stage, the code is incrementally modified to make the algorithm parallelizable. In the third stage, the programmer runs the translator and adjusts the code based on its recommendations to ensure error-free translation. At this point, a functional GPU version of the algorithm is available, and the process can be considered complete. If needed, in the fourth stage, the programmer refines the generated version by replacing individual computational kernels or virtual functions of the generated class.}\label{fig:concept1}
\end{figure}

Our approach, in some sense, "goes against the grain", as it is built on principles opposite to many existing ones:

\begin{enumerate} 
	\item We use an existing language (C++) as input without introducing new extensions (almost). Instead, we limit user capabilities to prevent code that cannot be fundamentally accelerated on massively parallel systems.
	
	\item The output is a low-level implementation (Vulkan, ISPC, etc.), with algorithmic optimizations applied directly to AST elements.
	
	\item The generated code is transparent ("white box") to the programmer, allowing for modifications that persist through subsequent re-generation (Fig. \ref{fig:concept1}). 
\end{enumerate}

Our programming technology employs pattern matching. These patterns do not define hardware functions or parallel constructs but encapsulate algorithmic and architectural knowledge. As a result, "clean code" that matches these patterns is inherently high-performance, with the translator, not the developer, responsible for hardware acceleration.

\FloatBarrier

\subsection{A no-API approach}

\begin{figure}[h]
	\centering
	\includegraphics[width=0.5\linewidth]{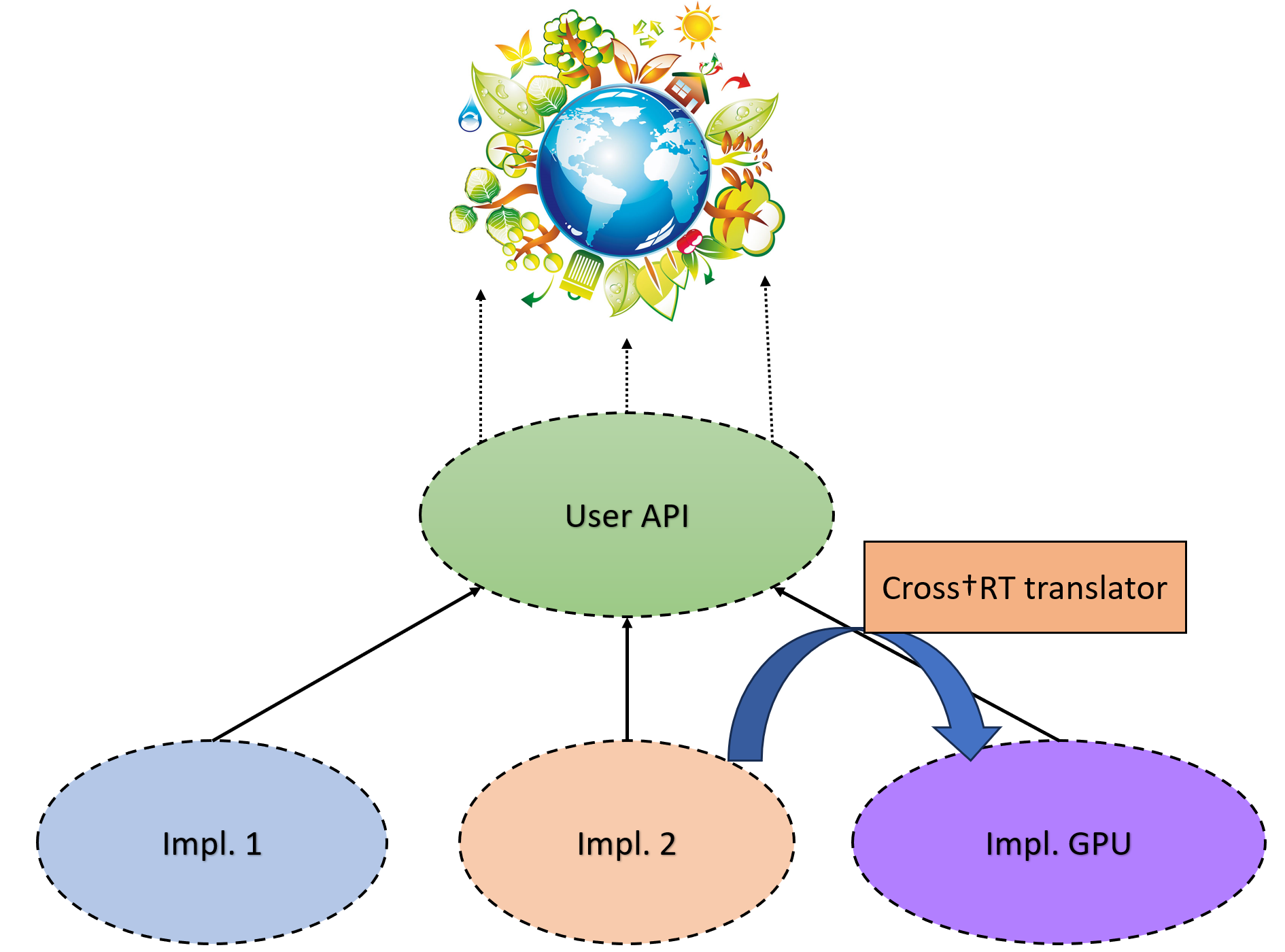}
	\caption{The generated class interacts with the external environment (the rest of the C++ program) through the API used by the base class (dashed arrows). In the generated code, these functions will be overridden. Therefore, in our programming technology, some virtual functions of the user class naturally serve as a boundary between the CPU and GPU address spaces. }\label{fig:concept2}
\end{figure}

Cross-platform compatibility is typically achieved by implementing an API on top of other APIs, a method used by most systems. For example, Halide \cite{halide} has it's own API. Our approach differs and is more close to Taichi\cite{taichi}. Our generator directly utilizes the API of the base class by generating derived class from the users one (fig. \ref{fig:concept2}).

\begin{figure}[h]
	\centering
	\begin{minipage}{0.49\textwidth}
		\centering
		\includegraphics[width=\textwidth]{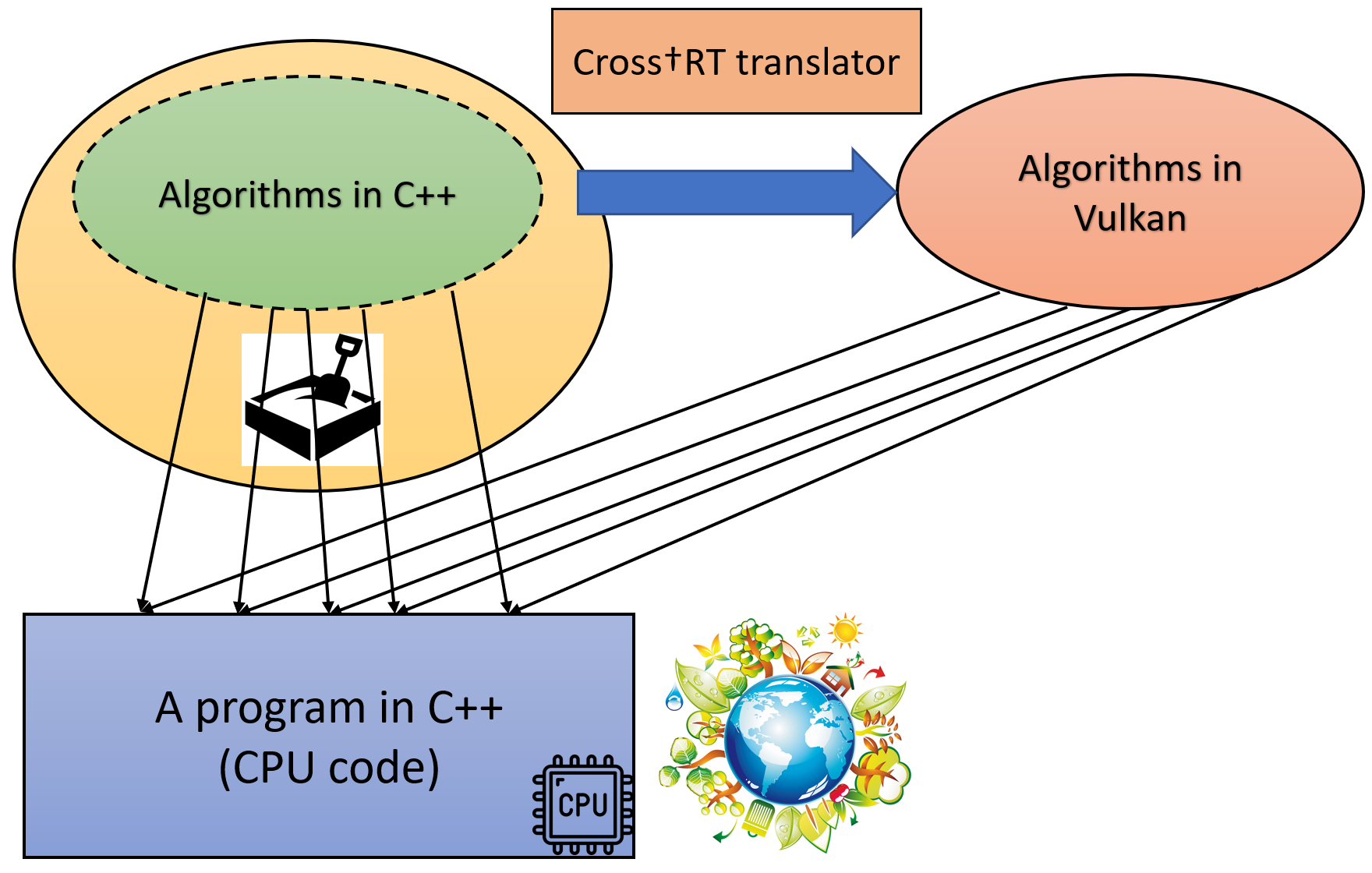}
	\end{minipage}
	\hfill
	\begin{minipage}{0.49\textwidth}
		\centering
		\includegraphics[width=\textwidth]{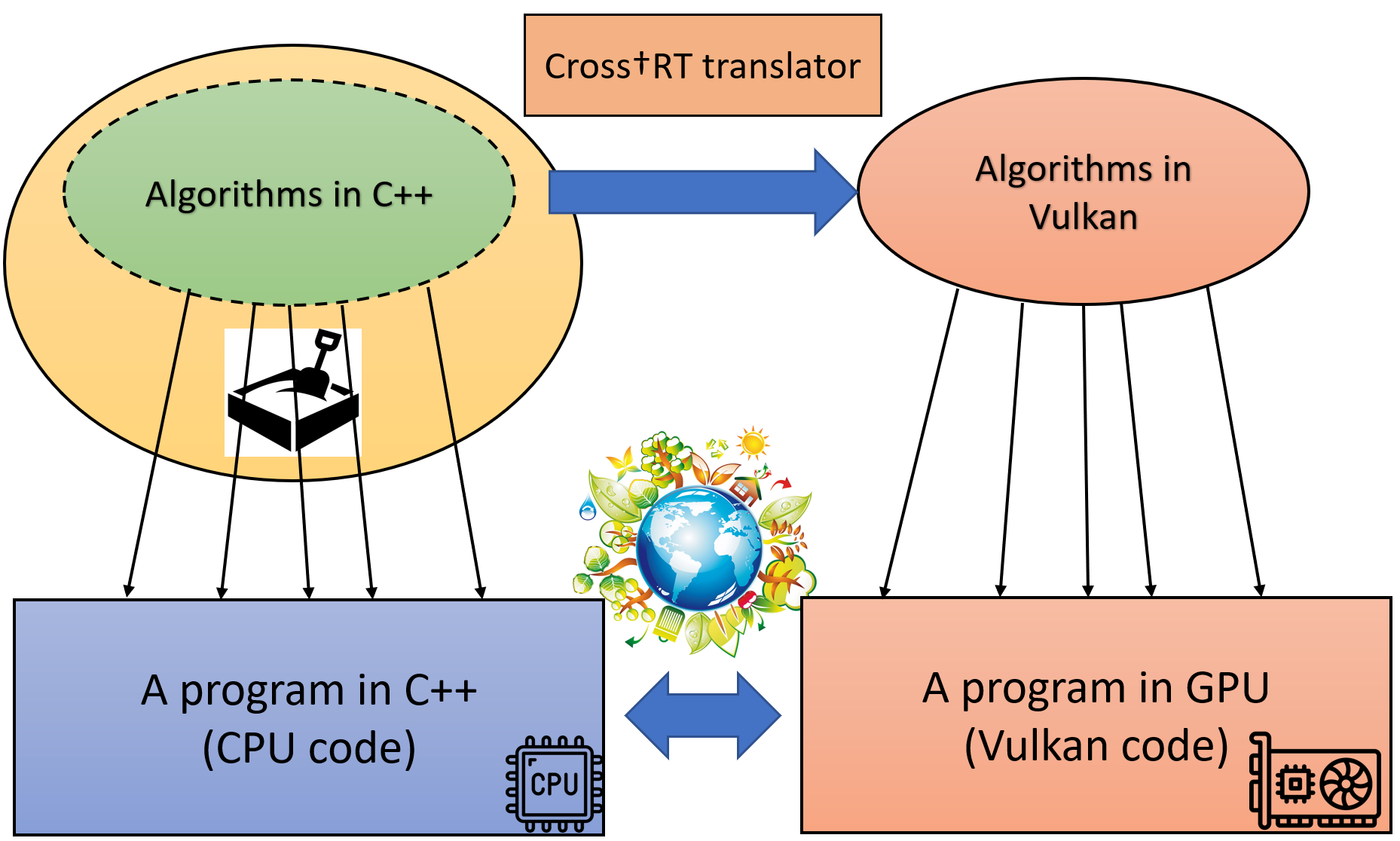}
	\end{minipage}
	\caption{Left: interaction of generated code with user-defined ordinary C++ code, right: interaction of generated code with user-defined Vulkan code. Both options are available. Black arrows denote virtual functions in the user class, which are overridden by our translator in the generated code. This allows the generated class to be seamlessly integrated into the user code without any modifications.}
	\label{fig:two_images}
\end{figure}
\FloatBarrier

\subsection{Translator Principles}

\begin{figure}[h]
	\includegraphics[width=\linewidth]{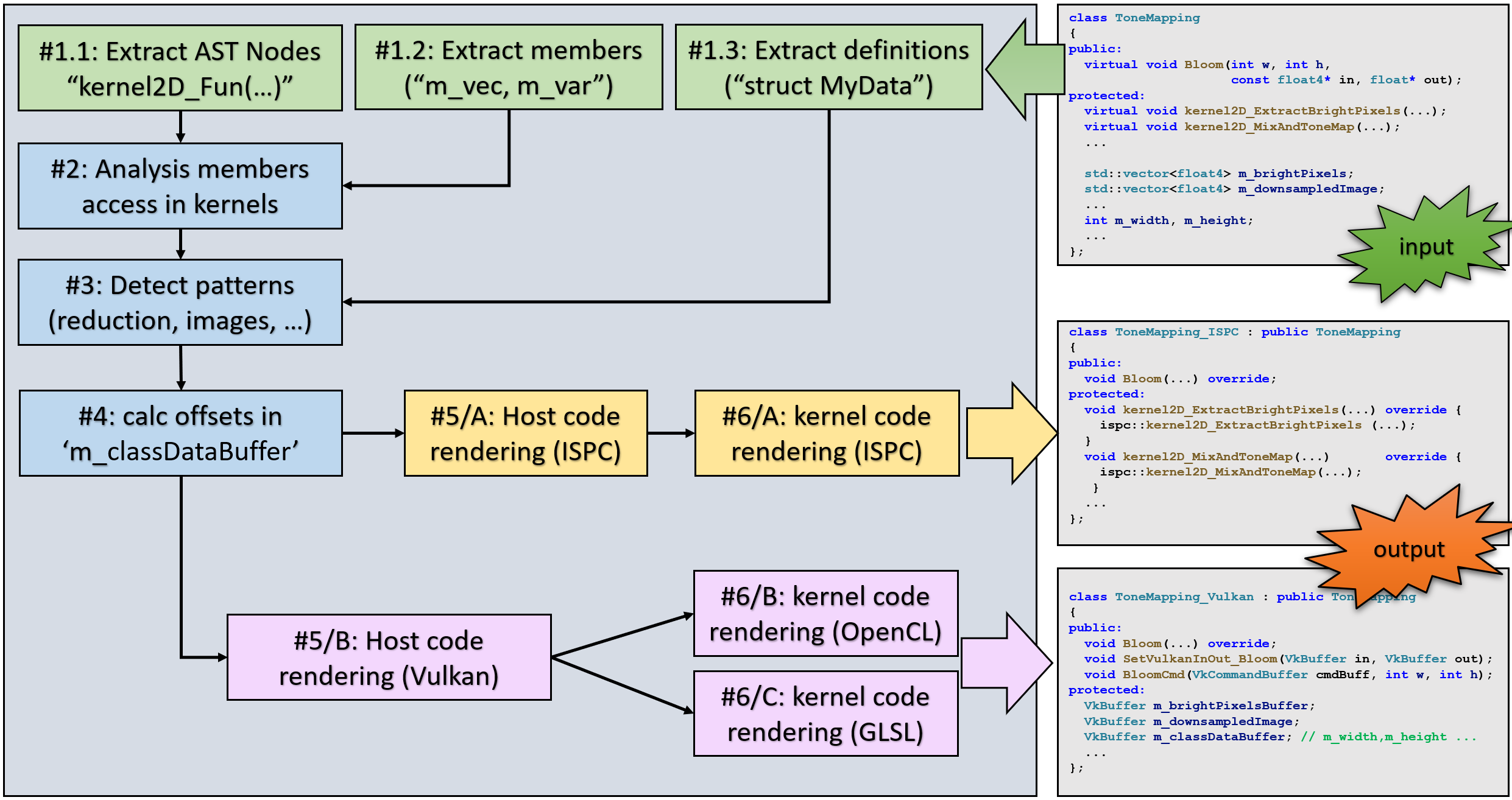}
	\caption{The flowchart of the proposed translator. Note that in the generated Vulkan code, the class includes member functions ("SetVulkanInOut\_Bloom" and "BloomCmd") designed to interact with existing data in GPU memory.}\label{fig:sceme}
\end{figure}

In our approach, the unit of "translation" is a class, whose name must be explicitly provided as input to the translator. From this, a derived class is generated, replacing the base implementation with a massively parallel version using the same algorithms (Fig. \ref{fig:sceme}). Integration of the generated class into existing code is achieved by simply replacing the pointer to the original class with a pointer to the generated class. The standard virtual function override mechanism then operates as usual.

\FloatBarrier

Let us consider the steps shown in Fig. \ref{fig:sceme}. In the first pass (green rectangles at the top), we extract AST nodes for \textit{kernels} and \textit{controller functions}, class member names and types, constants, and functions. Kernels are class member functions that contain nested loops and will be transformed into GPU compute kernels. Controller functions are member functions that invoke kernels and thus manage their execution. In the generated class, these functions will be replaced with code handling resource binding and kernel execution on the GPU.

In subsequent passes, we analyze the original kernel code (blue rectangles), examining which class members the kernels access and how this access occurs within loops. We also analyze other behaviors, specifically identifying parallel programming patterns in the sequential code: indirect kernel calls, reductions, virtual function calls, texture access, array appending, ray tracing, and the use of standard functionalities in controller functions such as memset, memcpy, sort, and scan. This is achieved through static analysis of the AST using Clang.

The last two steps in Fig. \ref{fig:sceme} involve rendering text templates via the “{{inja}}” library \cite{b49}, using different templates for ISPC (yellow rectangles) and Vulkan (purple rectangles). As a result, the output classes can be used as if they were manually implemented in Vulkan or ISPC.

\subsection{Design}

One of the key factors for achieving cross-platform compatibility, in our view, lies in the lightweight nature of the developed technologies. Solutions like CUDA and OptiX, aside from their proprietary nature, are extremely heavyweight and bring a large number of dependencies during installation. Unfortunately, working with modern parallel programming technologies makes it impossible to eliminate dependencies entirely. However, at the very least, these dependencies can be made modular based on the target platform for which the code is being built. The design of our programming technology strictly adheres to this concept.

When developing source code in C++, the developer only needs to use a few lightweight libraries, without additional dependencies. Our translator processes the user code, generating implementations for Vulkan or ISPC, which the user manually links to their program using CMake or any other build tool.

\subsection{Ecosystem}

Cross$^\dagger$RT ecosystem consists of source-to-source translator, which is based on kernel\_slicer \cite{kslicer}, and of several lightweight libraries, whose APIs are understood by our translator:

\begin{itemize}
	\item \textbf{LiteMath:} A header-only mathematical library for working with vectors and matrices in 2,000 lines of code. It is a lightweight alternative to the well-known glm library \cite{glm}. While our translator supports code using glm, in our projects, we aim for a more lightweight implementation, as glm's complexity often leads to difficulties in building applications with compilers other than the one used by its developers. Unlike glm, LiteMath does not use templates and is designed to be as simple and straightforward as possible.
	\item \textbf{LiteImage:} Template-based header in 300 lines of code which implements class for 2D images. Additional ``LiteImage.cpp'' file in 1500 lines of code has binding to popular stb\_image library for loading and saving images. 
	\item \textbf{LiteRT:} A header file (200 lines of code) containing the definition of \texttt{ISceneObject} interface. This interface handles adding geometric objects to the scene, instancing, and ray queries. It has implementations through Embree and a custom pure C++ implementation.
\end{itemize}

To utilize hardware acceleration (e.g., textures or ray tracing), the developer should use data types from these three provided libraries. It is important to note that the developer is not required to use our implementation; only the use of the libraries' interface is necessary. Additionally, when developing for the CPU, the user does not interact with our translator. The developed code is independent of it and can be used with any C++-compatible compiler or, for example, with CUDA, without involving our programming technology. Therefore, our translator does not introduce a dependency on itself, which is often a deterrent when choosing a programming technology. 

\subsection{Hardware Acceleration}

Our translator maps specific data types (C++ classes) and library functions to hardware-accelerated features on the GPU. For example, each class member of type \texttt{std::vector} is transformed into a \texttt{VkBuffer}. To add image support, users must employ a special templated \texttt{Image2D} class, which is a library implementation of images in C++. For hardware-accelerated ray tracing, users invoke member functions from a designated \texttt{ISceneObject} interface. On the CPU, this interface is implemented via Embree \cite{Embree}, while on the GPU, the translator generates GLSL code with hardware-accelerated ray tracing queries.

A few clarifications are necessary: first, the translator only targets the interfaces std::vector, Image2D, and ISceneObject — it does not concern itself with their CPU implementations, which can be entirely arbitrary. For instance, if a user cannot utilize Embree, they can create their own ISceneObject implementation. Second, for GPUs lacking specific hardware support (e.g., ray tracing), the translator can generate code with software implementations of the same functionality. To achieve this, we developed a mechanism to replace unsupported hardware features using a class composition pattern.

\subsection{Support for Older GPUs via composition pattern.}\label{classcomposition} Consider the three steps a programmer takes to replace unsupported hardware functionality of an interface (e.g., the ray tracing interface \texttt{ISceneObject}) with a custom algorithmic implementation, which will be similarly translated to the GPU by our compiler as other user code:

\begin{enumerate} 
\item In the first step, the programmer uses the "composition" programming pattern and includes a pointer to the \texttt{ISceneObject} interface within their \texttt{MyRender} class. 
	
\item In the second step, the programmer must create a custom implementation of \newline \texttt{ISceneObject} (called \texttt{MyRayTracer}) in C++ and debug it on the CPU. 

\item In the third step, the programmer explicitly instructs the compiler via command-line arguments to use their custom \texttt{MyRayTracer} class for implementing the \texttt{ISceneObject} interface in the \texttt{MyRender} class. 
\end{enumerate}

Our translator merges together \texttt{MyRender} and \texttt{MyRayTracer} as if all data and functions of \texttt{MyRayTracer} are declared within \texttt{MyRender}. To avoid name conflicts, renaming is employed: a prefix \texttt{m\_pRayTracer\_} is added to all data and functions of the \texttt{MyRayTracer} class (listing \ref{composition}).

{\begin{lstlisting}[language=C++, basicstyle=\small]
struct MyRender {
  void kernel2D_DoSomeRayTrace(...) { 
    auto hit = m_pRayTracer->RayQuery_NearestHit(...);  
  }
  std::shared_ptr<ISceneObject> m_pRayTracer;  
};
\end{lstlisting}}
{\captionof{lstlisting}{An example of class composition pattern \label{composition}}}

\subsection{Support for future GPUs} Given the vast array of hardware accelerations, it is infeasible to support all of them in our translator (the same can be said for any other existing programming technology due to almost infinite amount of work). However we aim to incorporate future hardware acceleration features not yet available in current GPUs. To address this, users can create a derived class from the class generated by our translator. In this derived class, users can make necessary modifications by directly interacting with Vulkan or ISPC, replacing specific kernels and virtual functions. Thus, user modifications persist even when the source code is changed and the GPU implementation class is regenerated.

\subsection{Patterns and Their Detection}

For us, a pattern is any set of frequently used language constructs and/or library calls in a program that must be specially implemented on a GPU (or other massively parallel computing system). Here are some key patterns detected by our translator:

\begin{enumerate} 
	
\item Parallel reduction, identified by detecting specific access patterns to class variables; \item Parallel prefix sum and sorting, appearing as calls to library functions like \newline  \texttt{std::inclusive\_scan}, \texttt{std::exclusive\_scan}, and \texttt{std::sort} in control functions; 
\item Calls to specific library interfaces: texture access, ray tracing queries, texture array access, bit representation conversions, and complex number operations; 
\item Virtual function calls; 
\item Various uses of pointers: For example, the pointer addition operation in the function argument $f(vec1.data() + offset)$ is a pattern that we handle separately to support limited pointer functionality which is absent in GLSL by default. 
\end{enumerate}

\subsection{Virtual functions}

Cross$^\dagger$RT implements a four-tier scheme for virtual functions.

\textbf{Level Zero.} This is complete de-virtualization, similar to the strategy used in Luisa. However, unlike Luisa, Cross$^\dagger$RT accomplishes de-virtualization not through templates, but via a class composition mechanism discussed in section \ref{classcomposition}. The primary concept here is that the translator explicitly receives instructions that a certain pointer (in the considered example, this is m\_pRayTracer) will have a specific type. Thus, the generated code can work directly with this specific type using either dynamic\_cast (in CPU code) or directly with the class data (in GPU code and ISPC kernels). For the considered example, a class MyRender\_BVHRT\_GPU would be generated, in which the members of the MyRender and BVHRT classes are combined into a single namespace using renaming for the members of the BVHRT class. A downside of this approach emerges when the base class uses a multitude of abstract pointers with various possible hierarchies, leading to a combinatorial explosion. Thus, it's impractical to use this approach if there are more than 2-3 such pointers. Although what we perform for devirtualization at level zero closely resembles templates (for example, \texttt{MyRender\_BVHRT\_GPU} could be interpreted as a C++ template specialization for \texttt{MyRender<BVHRT>\_GPU} or \texttt{MyRender<BVHRT,GPU>}), we intentionally avoid using templates for a specific reason: we do not wish to complicate the user code, which otherwise would have to explicitly convert \texttt{MyRender} into a template class.

\textbf{Level One.} At this level, we implement a simplified version of dynamic dispatch in the compute shader using a switch construction. We allow the user to have an array (of type \texttt{std::vector}) of abstract pointers. However, at this level, we require that all implementations possess an identical set of data that must be defined in the interface. This requirement stems from a GLSL limitation, where pointers are absent and there is no capability to perform pointer casting based on dynamic type information. And thus all derived classes should use same structure.

\textbf{Level Two.} At this level, we allow the user to extend derived classes with new data. All objects must be created on the CPU before the special function \texttt{CommitDeviceData()} is called, which in the generated class reorders objects by data type and places them in different memory regions on the GPU so that all objects of the same type are stored sequentially in memory. Subsequently, GLSL code uses the EXT\_Buffer\_Reference extension to access a specific memory region on the GPU based on the object's internal index within its type. We explicitly separate the first and second levels since not all GPUs support EXT\_Buffer\_Reference.

\textbf{Level Three.} This level differs from the previous one only in that instead of compute shader and switch construct it generates a RayGen shader in ray tracing pipeline utilizing genuine virtual function mechanism through callable shaders.

\subsection{Mapping to intersection shader}

The support for intersection shaders in our technology is implemented as an extension of virtual function. Instead of callable shaders, code is generated featuring an intersection shader that is invoked upon hitting a BVH tree leaf. To utilize hardware traversal of the BVH tree, the user must implement a function RayQuery\_NearestHit within their class (e.g., BVHRT), where at a certain point, the virtual function Intersect must be called (listing \ref{intersection}). The Intersect function has a set of fixed parameters (ray, pointer to a CRT\_Hit structure, and a CRT\_LeafInfo structure) but can also accept additional parameters if necessary. For instance, a pointer to the BVHRT class can be passed inside the Intersect function, which is useful for accessing data stored within Intersect. In the absence of ray-tracing pipeline support, the user can generate a compute shader, where a switch statement will be generated for Intersect according to the second or first level of virtual function support.

{\begin{lstlisting}[language=C++, basicstyle=\small]
CRT_Hit BVHRT::RayQuery_NearestHit(float4 rPosAndNear, float4 rDirAndFar)
{
  CRT_Hit res = {...};
  // any custom traveral code here
  // for example [Aila and Laine 2009] while-while traversal
  while (stack is not empty) {
    while(searching for Leaf) {
      // traversal code here
      ...
    }

    CRT_LeafInfo leafInfo;
    leafInfo.geomId = geomId;
    leafInfo.instId = curr_instance_id;
    leafInfo.aabbId = curr_leaf_aabb_id;
    ...
    m_geom[geomId]->Intersect(rPosAndNear,rDirAndFar,leafInfo, 
                              &res, ... );  
    // pop nodes from stack and continue traversal
  }	
  return res;
}
\end{lstlisting}}
{\captionof{lstlisting}{An example of class composition pattern \label{intersection}}}

Thus, supporting hardware functionality does not require introducing a special software model or creating new language constructs. The implementation through either a compute shader or a ray-tracing pipeline is a translator option and is not explicitly expressed in the algorithm that describes BVH tree traversal and intersection search. We have validated this functionality on SDF functions (section \ref{sdftrace}) and radiance fields (section \ref{radiancefields}).

%\subsection{Specialization constants}
%security and safety issue: are spec constants defined in VulkanSC ?

%%%%%%%%%%%%%%%%%%%%%%%%%%%%%%%%%%%%%%%%%%
\section{Validation and results}

In our view, the validation of a programming technology should meet two criteria. First, it must be tested on tasks for which the technology was specifically developed. This demonstrates the flexibility and capability of the technology in addressing its intended problem domain. Second, validation should not be limited to toy examples (which were already validated, for example, in \cite{kslicer} for kernel\_slicer) but should also include tasks that closely resemble real-world challenges and applications. Selected scenarios represent complex programs that solve specific tasks comprehensively. For example, path tracing includes implementations of various material models, while BVH construction involves non-trivial parallel hierarchy generation algorithms. Thus, when comparing the implemented algorithms to alternatives, we are actually comparing problem-solving approaches using different programming methodologies, not just evaluating compiler efficiency on similar codebases. The latter, in our view, is of limited value, akin to comparing CUDA with SPIR-V/GLSL/Vulkan compiler on the Mandelbrot fractal evaluation. While performance differences between CUDA and GLSL may exist, this is not the focus of our comparison.

\subsection{LBVH construction and traversal}

Building a BVH tree in parallel on a GPU is a challenging task and serves as an effective stress test for the programming technology. Tree construction is a complex and inefficient task for GPUs, primarily limited by memory bandwidth. In contrast, tree traversal during ray tracing is easily parallelizable and accelerates well (although it is also often constrained by memory bandwidth).  Unlike previous cases, for Cross$^\dagger$RT, we did not introduce any new functionality compared to kernel\_slicer. We utilized features of kernel\_slicer such as code generation for sorting and prefix sum, as well as indirect dispatch. We implemented the parallel algorithm for constructing an LBVH tree as described in the paper by Tero Karras \cite{Karras12} and compared this implementation to the LBVH tree construction in Embree \cite{Embree} and HIPRT \cite{hiprt} for construction  (tables \ref{tab:rtx2070}, \ref{tab:adreno730}, \ref{tab:rtx2070hippie}), and traversal (tables \ref{tab:rtx2070trace}, \ref{tab:adreno730trace}, \ref{tab:rtx2070tracehippie}).

The tree construction algorithm based on Karras' approach involves several parallelizable steps:

\begin{enumerate} 
\item \textbf{Morton code \cite{Morton66} calculation.} A 3D spatial index is computed for each primitive. This requires a parallel reduction to find the AABB that encloses all primitives, followed by Morton code computation.
	
\item \textbf{Sorting Morton codes.} Key-value pairs are sorted using bitonic or radix sort, where the key is the Morton code and the value is the object's original index.
	
\item \textbf{Leaf node generation.} Primitives with the same Morton code are grouped into leaf nodes using a parallel prefix sum. Bounding boxes for these nodes are also computed.
	
\item \textbf{Hierarchy construction} according to Karras algorithm \cite{Karras12}. Using prefix codes, tree nodes are built in parallel, ensuring that nodes with distant Morton codes are combined only at higher levels for better tree quality.
	
\item \textbf{Bounding box refitting.} The tree is traversed bottom-up to refit bounding boxes. The node order complicates this, requiring multiple parallel passes, but a single-pass solution is available in \cite{fastbuild5}. 
\end{enumerate}

Our implementation is shown in listing \ref{list:constrolfunc}. Functions like \textit{std::exclusive\_scan} and \textit{std::sort} with lambda expressions are recognized by our compiler. It generates bitonic sort and parallel prefix sum implementations in Vulkan for these functions. In the ISPC backend, these functions remain unchanged, with only the code inside the kernels being replaced.

{\begin{lstlisting}[language=C++, basicstyle=\small]
void LBVH_Karras::BuildFromBoxes(const Box4f* in_boxes,uint a_boxCount, 
	                         BVHNode* out_tree) 
{
  // (0) input => (bboxMin, bboxMax)
  kernel1D_Reduction(a_boxes, a_boxCount);
  // (1) input => runCodesAndIndices
  kernel1D_EvalMC(a_boxes, a_boxCount, runCodesAndIndices.data());  
  // (2) sort runCodesAndIndices by morton code
  std::sort(runCodesAndIndices.begin(), runCodesAndIndices.end(), 
            [](uint2 a, uint2 b) { return a.x < b.x; }); 
  // (3) make leafes array
  // runCodesAndIndices => (codesEq, prefixCodeEq)
  kernel1D_AppendInit(runCodesAndIndices.data(), a_boxCount, 
                      codesEq.data(), prefixCodeEq.data());                   
  std::exclusive_scan(prefixCodeEq.begin(), prefixCodeEq.end(), 
                      prefixCodeEq.begin(),0);                     
  // (codesEq, prefixCodeEq) => (newIndices, compressedCodes)                               	  
  kernel1D_AppendComplete(codesEq.data(),  a_boxCount,
                          prefixCodeEq.data()); 
  // a_boxes => a_outNodes
  kernel1D_MakeLeaves(a_boxes, a_boxCount, a_outNodes);                     
  // (4) Karras algorithm from his paper                         
  kernel1D_Karras12(out_tree);  
  // (5) Refit       
  // a_outNodes => runCodesAndIndices (used as temp buffer)               
  kernel1D_RefitInit(a_outNodes, runCodesAndIndices.data()); 
  // (a_outNodes,runCodesAndIndices) => 
  // (a_outNodes,runCodesAndIndices)
  for(int pass = 0; pass < 32; pass++)
    kernel1D_RefitPass(a_outNodes, runCodesAndIndices.data());
}			
	
\end{lstlisting}
{\captionof{lstlisting}{The control function responsible for launching compute kernels and implementing the tree construction using the Karras algorithm.}\label{list:constrolfunc}}}

Let us highlight several features of the proposed programming technology that we encountered during the implementation of the parallel tree construction algorithm.

\begin{enumerate} 
\item The original tree construction code is written in standard C++11 and can be compiled by any compiler that supports this language (with minor modifications, it could even be compiled with C++98). 
\item We maintained a relatively clean, single-threaded description of the algorithm, without using any parallel programming constructs or directives. Although our translator does not prohibit adding such directives (e.g., OpenMP). 
\item Nevertheless, we had to parallelize the tree construction algorithm in several places: this includes parallel insertion of elements into a buffer, which we explicitly implemented using prefix sum (listing \ref{list:constrolfunc}, \textit{std::exclusive\_scan} call), the Karras algorithm \cite{Karras12} that we used to build the tree hierarchy, and finally, a parallel ``refit'' which compute bounding boxes of nodes in a bottom-up manner. 
\end{enumerate}

\subsubsection{Comparison with Embree}

Embree a solid reference point for comparisons with CPU-based implementations as it has long development history and is widely adopted in the industry. We conducted a comparison with this solution on both desktop and mobile platforms to demonstrate that the cross-platform nature of our proposed solution enables the use of a mobile graphics processor, whereas this is not possible with Embree. Our comparison shows that due to the capabilities of the programming technology and hardware, we can achieve better results even with a less optimized original implementation of the tree construction or traversal algorithm. 

{\begin{table}[h]
		\begin{tabular}{|c|c|c|c|c|c|}
			\hline
			Scene & N Primitives & Embree & Ours (CPU) & Ours (GPU) & speed-up \\
			\hline
			\hline
			teapot     & 25K  & 6.67 ms & 5.48 ms & \textbf{1.41 ms} & 3.9 \\
			\hline
			sponza     & 66K  & 7.48 ms & 9.71 ms & \textbf{2.47 ms} & 3.9 \\
			\hline
			bunny      & 144K & 13.5 ms & 21.6 ms & \textbf{4.54 ms} & 4.8 \\ 
			\hline
			cry-sponza & 243K & 14.6 ms & 30.0 ms & \textbf{4.80 ms} & 6.2 \\ 
			\hline
			dragon     & 871K & 61.7 ms & 149 ms & \textbf{20.2 ms} & 7.3 \\ 
			\hline
			car        & 1.6M & 88.0 ms & 199 ms & \textbf{35.4 ms} & 5.6 \\ 
			\hline
		\end{tabular}
	\caption{\label{tab:rtx2070} Comparison with Embree. \textbf{Construction} time on a desktop system. Hardware: Intel Core i7-9700K. 3.60GHz (8 cores); GPU: Nvidia RTX2070. Ubuntu Linux 23.}
	\end{table}

{\begin{table}[h]
		\begin{tabular}{|c|c|c|c|c|c|}
			\hline
			Scene & Primitives & Embree & Ours (CPU) & Ours (GPU) & speed-up \\
			\hline
			\hline
			teapot     & 25K  & 9 ms & 7.7 ms & \textbf{4.6 ms} & 1.6 \\
			\hline
			sponza     & 66K  & 13.9 ms & \textbf{11.8 ms} & 14.9 ms & 0.8 \\
			\hline
			bunny      & 144K & \textbf{31.3 ms} & 34.0 ms & 32.4 ms & 1.1 \\ 
			\hline
			cry-sponza & 243K & 40.71 ms & 42.8 ms & \textbf{35.0 ms} & 1.2 \\ 
			\hline
			dragon     & 871K & 120 ms & 165 ms & \textbf{109 ms} & 1.5 \\ 
			\hline
			car        & 1.6M & 265 ms & 379 ms & \textbf{210 ms} & 1.8 \\ 
			\hline
		\end{tabular}
		\caption{\label{tab:adreno730} Comparison with Embree. \textbf{Construction} time on a mobile system. Hardware: CPU - Qualcomm Snapdragon 8 Gen 1; GPU - Adreno 730. OS: Android 13 Tiramisu. }
	\end{table}

{\begin{table}[h]
		\begin{tabular}{|c|c|c|c|c|c|}
			\hline
			Scene & Primitives & Embree & Ours (CPU) & Ours (GPU) & Ours (RTX) \\
			\hline
			\hline
			teapot     & 25K  & 84 ms & 123 ms & 2.4 ms & \textbf{2.0 ms} \\
			\hline
			sponza     & 66K  & 126 ms & 365 ms & 7.0 ms & \textbf{2.5 ms} \\
			\hline
			bunny      & 144K & 87 ms  & 161 ms & 2.71 ms & \textbf{2.1 ms} \\ 
			\hline
			cry-sponza & 243K & 177 ms & 486 ms & 14.3 ms & \textbf{3.4 ms} \\ 
			\hline
			dragon     & 871K & 138 ms & 186 ms & 4.7 ms & \textbf{1.9 ms} \\ 
			\hline
			car        & 1.6M & 118 ms & 319 ms & 8.6 ms & \textbf{2.4 ms} \\ 
			\hline
		\end{tabular}
		\caption{\label{tab:rtx2070trace} Comparison with Embree. \textbf{Traversal} time for approximately 4 million (2048x2048) primary rays on a mobile system for a tree constructed using the LBVH algorithm. Hardware: Intel Core i7-9700K. 3.60GHz (8 cores); GPU: Nvidia RTX2070. Ubuntu Linux 23.}
	\end{table}

{\begin{table}[h]
		\begin{tabular}{|c|c|c|c|c|}
			\hline
			Scene & Primitives & Embree & Ours (CPU) & Ours (GPU) \\
			\hline
			\hline
			teapot     & 25K  & 159 ms & 202 ms & \textbf{13.0 ms} \\
			\hline
			sponza     & 66K  & 223 ms & 490 ms & \textbf{33.8 ms} \\
			\hline
			bunny      & 144K & 135 ms & 207 ms & \textbf{15.8 ms}  \\ 
			\hline
			cry-sponza & 243K & 357 ms & 860 ms & \textbf{69.0 ms} \\ 
			\hline
			dragon     & 871K & 163 ms & 215 ms & \textbf{44.0 ms}  \\ 
			\hline
			car        & 1.6M & 212 ms & 527 ms & \textbf{48.9 ms} \\ 
			\hline
		\end{tabular}
	\caption{\label{tab:adreno730trace} Comparison with Embree. Tree traversal time for approximately 4 million (2048x2048) primary rays on a mobile system for a tree constructed using the LBVH algorithm. Hardware: CPU - Qualcomm Snapdragon 8 Gen 1 (arm64, 8 cores); GPU - Adreno 730. OS: Android 13 Tiramisu.}
	\end{table}
    \FloatBarrier

\subsubsection{Comparison with HIPRT}

Let us focus on the tree construction speed first. Our GPU-based tree construction, automatically derived from C++ code, achieves approximately the same speed as the manually optimized CUDA/HIP implementation from HIPRT \cite{hiprt} (Table \ref{tab:rtx2070hippie}). The slight difference in construction speed is most likely caused by the fact that HIPRT employs a more efficient radix sort algorithm with a complexity of $O(Nk)$, whereas implementation from kernel\_slicer uses bitonic sort with a complexity of $O(N\log^2(N))$. Additionally, one of the main drawbacks of bitonic sort is the need to increase the size of the sorted array to the nearest power of two. For example, when sorting 243 thousand primitives (cry-sponza), we sort 256 thousand primitives, and for 280 thousand primitives (conference), we must sort 512 thousand elements. Similarly, for 66 thousand primitives (sponza), we sort 128 thousand elements, which explains the performance drop in the sponza scene (Tables \ref{tab:rtx2070}, \ref{tab:adreno730}).

{\begin{table}[!h]
		\begin{tabular}{|c|c|c|c|}
			\hline
			Scene & Primitives & HIPRT & Ours (GPU) \\
			\hline
			\hline
			teapot     & 25K  & 2.2 ms & \textbf{1.4 ms} \\
			\hline
			conference & 144K & \textbf{3.6 ms} & 4.2 ms \\ 
			\hline
			cry-sponza & 243K & \textbf{4.5 ms} & 4.8 ms \\ 
			\hline
			dragon     & 871K & \textbf{12 ms} & 20 ms \\ 
			\hline
		\end{tabular}
	\caption{\label{tab:rtx2070hippie} Comparison with HIPRT. \textbf{Construction} time on a stationary system using the LBVH algorithm. Hardware: CPU - Intel Core i7-9700K 3.60GHz (8 cores); GPU - Nvidia RTX2070. OS: Ubuntu Linux 23.}
	\end{table}

{\begin{table}[h]
		\begin{tabular}{|c|c|c|c|c|}
			\hline
			Scene & Primitives & HIPRT & Ours (GPU) & Ours (RTX) \\
			\hline
			\hline
			teapot     & 25K  & 2.6 ms  & 2.4 ms  & \textbf{2.0 ms} \\
			\hline
			conference & 280K & 6.00 ms & 6.20 ms & \textbf{3.0 ms} \\ 
			\hline
			cry-sponza & 243K & 11.2 ms & 14.3 ms & \textbf{3.4 ms} \\ 
			\hline
			dragon     & 871K & 4.49 ms & 4.70 ms & \textbf{1.9 ms} \\ 
			\hline
		\end{tabular}
	\caption{\label{tab:rtx2070tracehippie} Comparison with HIPRT. Tree traversal time for approximately 4 million (2048x2048) primary rays on a desktop. For a tree constructed using the LBVH algorithm. Hardware: GPU - Nvidia RTX2070. OS: Ubuntu Linux 23. Column ``Ours (RTX)'' is hardware accelerated ray tracing generated by our translator. We provide this timing here to assess the efficiency of the software implementation. }
	\end{table}

\begin{figure}[h]
	\centering
	\begin{tabular}{ccc}
		\includegraphics[width=0.3\textwidth]{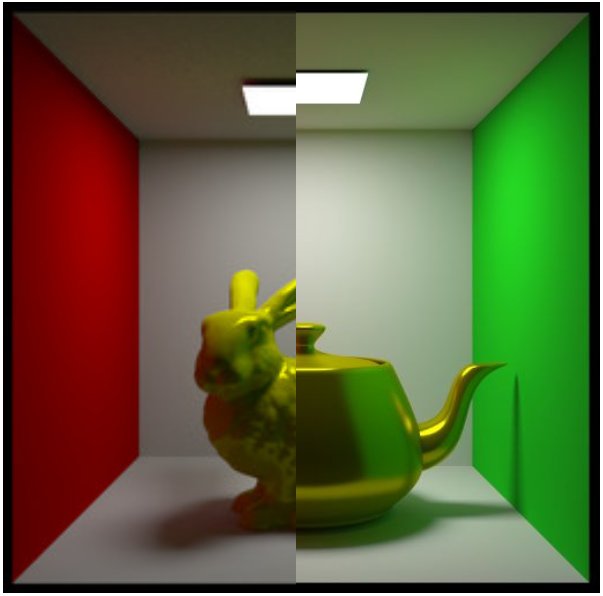} & \includegraphics[width=0.3\textwidth]{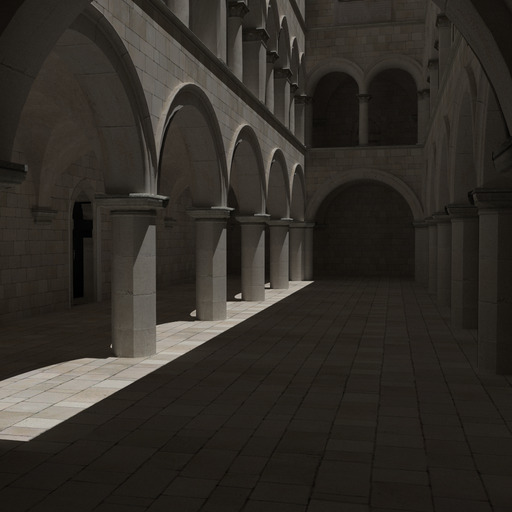} & \includegraphics[width=0.3\textwidth]{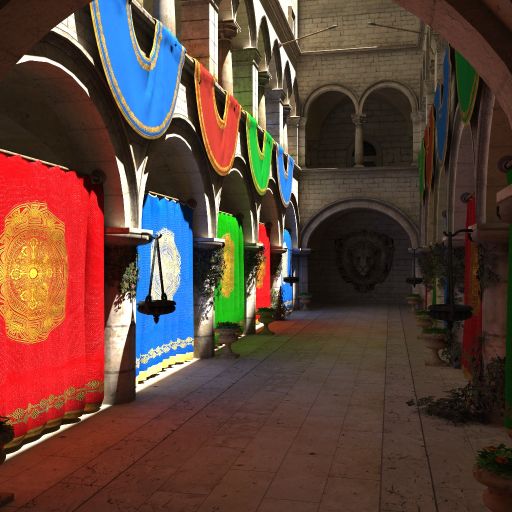} \\
		bunny/teapot & sponza & cry-sponza \\
		\includegraphics[width=0.3\textwidth]{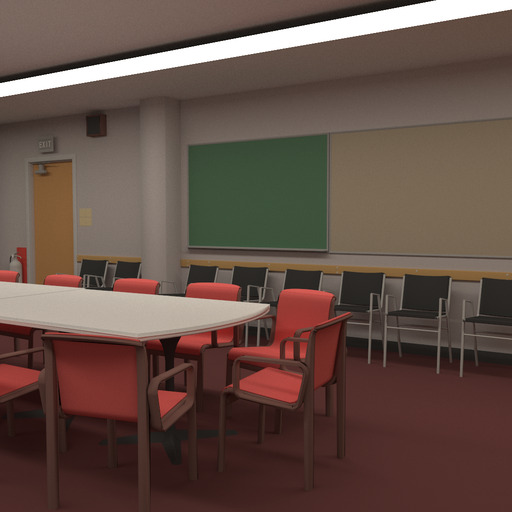} & \includegraphics[width=0.3\textwidth]{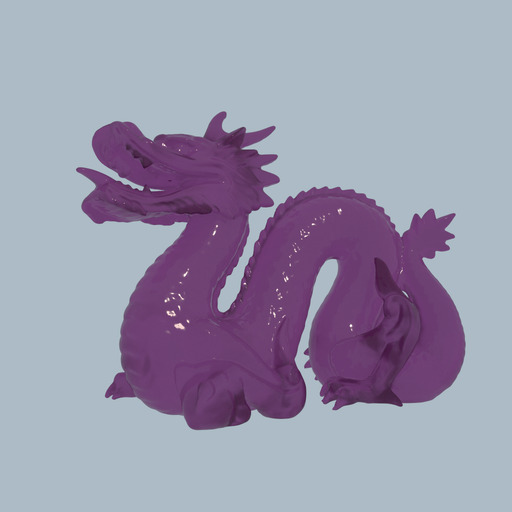} & \includegraphics[width=0.3\textwidth]{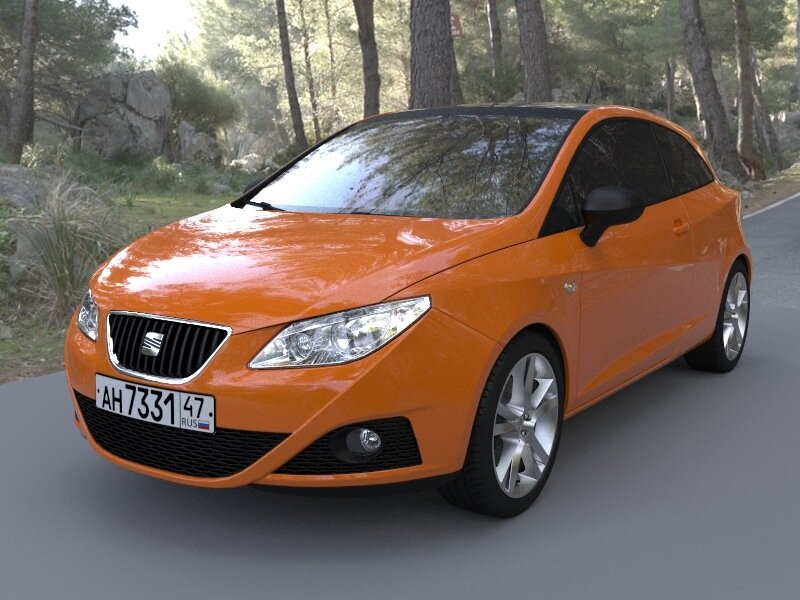} \\
		conference & dragon & car \\
	\end{tabular}
	\caption{Images of the test scenes from the viewpoint used in the LBVH tree construction and traversal. The images themselves were generated using Monte Carlo Path Tracing.}\label{fig:scenes}
\end{figure}
\FloatBarrier

For the desktop GPU (Table  \ref{tab:rtx2070}) we achieve a speedup of 4--7 times compared to the parallel CPU version, which roughly correlates with the difference in memory bandwidth between the CPU and GPU. On the mobile GPU (Table \ref{tab:adreno730}), significant speedup is not achieved, which we attribute to the weaker memory subsystem of mobile GPUs and, as previously discussed, a certain loss of efficiency in sorting.

The situation changes for tree traversal (ray tracing). Here, there are a substantial number of arithmetic operations. As a result, compared to the CPU implementation, we can achieve speedups of up to 40 times for software and up to 130 times for hardware implementation on the GPU for primary rays in a desktop system and 10-15 times on a mobile GPU.

\subsection{Signed Distance Functions}\label{sdftrace}

We have implemented several algorithms for tracing SDF functions from \cite{SDF22}, specifically the Sparse Brick Set (SBS) and Sparse Voxel Set (SVS). We measured the performance of ray tracing and path tracing for different model sizes (fig. \ref{fig:sdf_models}, fig. \ref{fig:sdf_lods}), as the key issue in SDF representations such as SBS and SVS is the tree depth. It can significantly affect rendering time, and we aimed to analyze the gains from hardware acceleration.

\begin{figure}[H]
	\centering
	\includegraphics[width=0.5\linewidth]{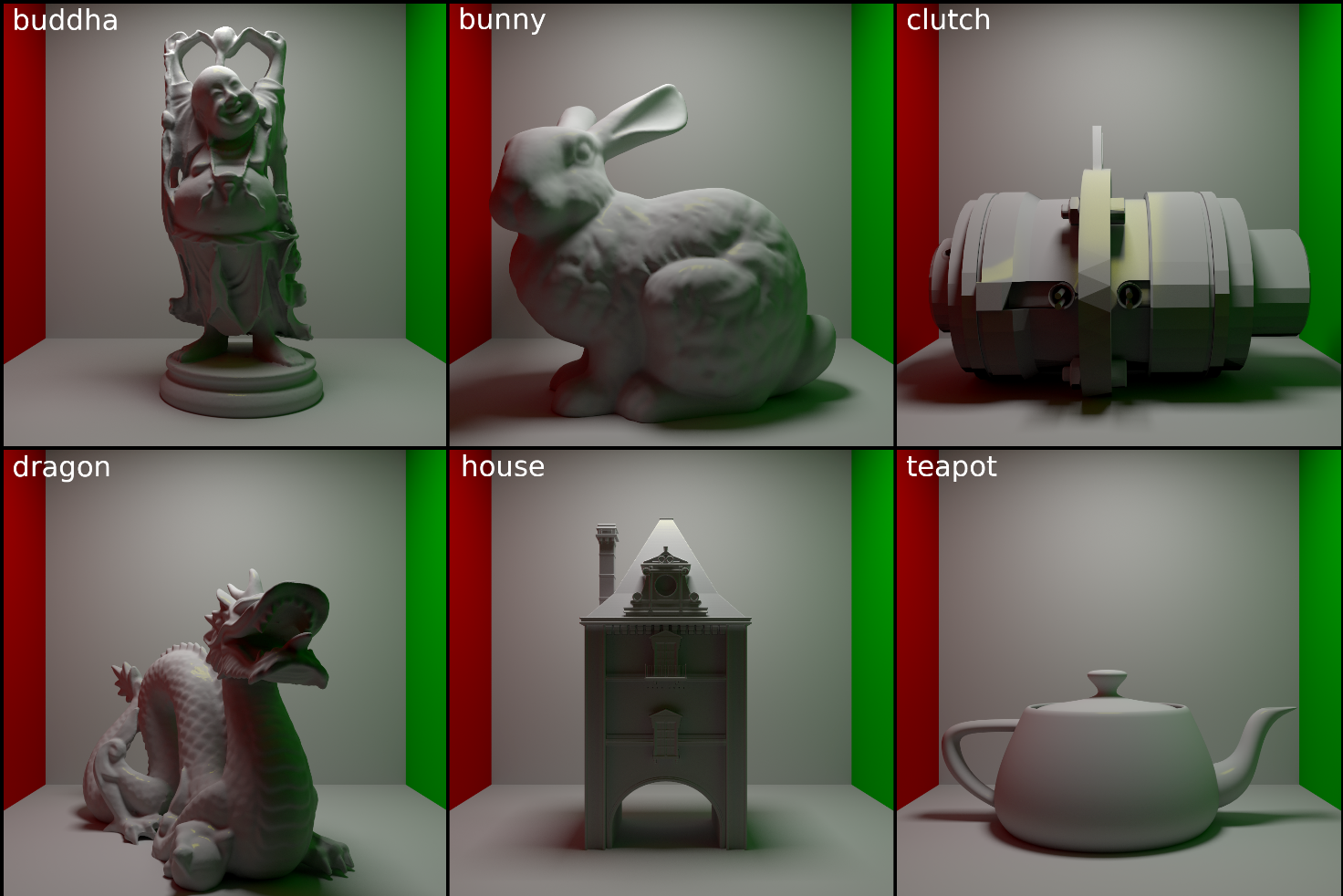}
	\caption{Examples of models used in performance analysis.\label{fig:sdf_models}}
\end{figure}

\begin{figure}[H]
	\includegraphics[width=\linewidth]{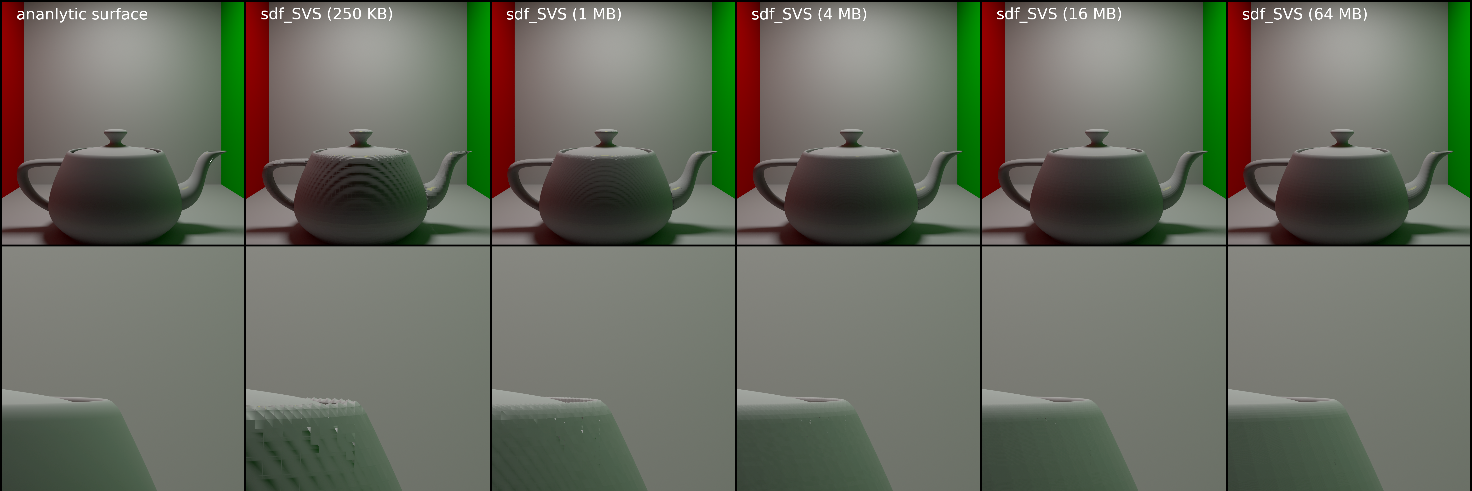}
	\caption{A teapot model represented as SVS with different model size.\label{fig:sdf_lods}}
\end{figure}

\begin{figure}[H]
	\includegraphics[width=\linewidth]{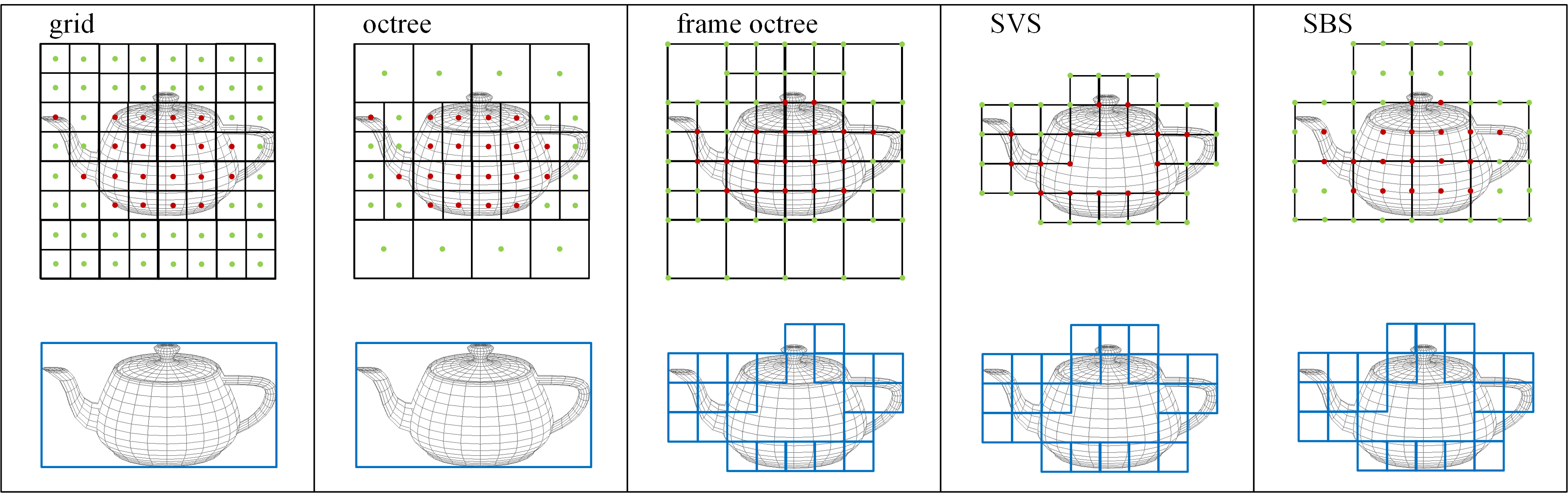}
	\caption{SDF representations used in performance analysis. From left to right: grid, octree, octree with values stored in corners (frame\_octree), sparse voxel set (SVS), sparse brick set (SBS). The upper diagrams demonstrate the partitioning of the space for each of the representations. The lower diagrams show the BVH leaf nodes for each of the representations. The green and red dots are the positions, where signed distance is saved. Green ones are outside the object, red ones - inside. The lower diagrams show non-empty (blue) BVH leaf nodes for each of the views. \label{fig:sdf_types}}
\end{figure}

An octree with corner values (\textbf{frame\_octree}). Saves 8 distances in the corners of each leaf node. This approach allows to calculate distances more accurately and significantly increases the locality of access to memory, however, it requires a significant amount of memory, since the values are duplicated between all voxels sharing sides.

\textbf{Sparse Voxel Set (SVS).} This representation consists of a set of voxels arranged as small regular grids within the leaves of a BVH tree. In SVS, each voxel stores 8 distances at its corners, and a BVH tree is built over the set of voxels, enabling efficient voxel traversal during ray tracing, including with hardware acceleration \cite{SDF22}. Since each voxel has a bounding box, the values inside the voxels can define distances relative to the voxel size, allowing for a more compact representation (8-bit in \cite{SDF22}).

\textbf{Sparse Brick Set (SBS).} This approach is similar to SVS but differs in that a small regular grid (e.g., 2x2x2 or 4x4x4) is stored within the voxels. This reduces the duplication of distance values compared to SVS but requires more computation during intersection tests. 

SDF grid was also included into the analysis as a simple baseline, while octree was discarded after the first few experiments that shown it low performance (two orders of magnitude slower rendering than other representations). To calculate the intersection of a ray and a surface, we employed the Newton method method proposed in \cite{SDF22}.

\begin{figure}[h]
	\includegraphics[width=\linewidth]{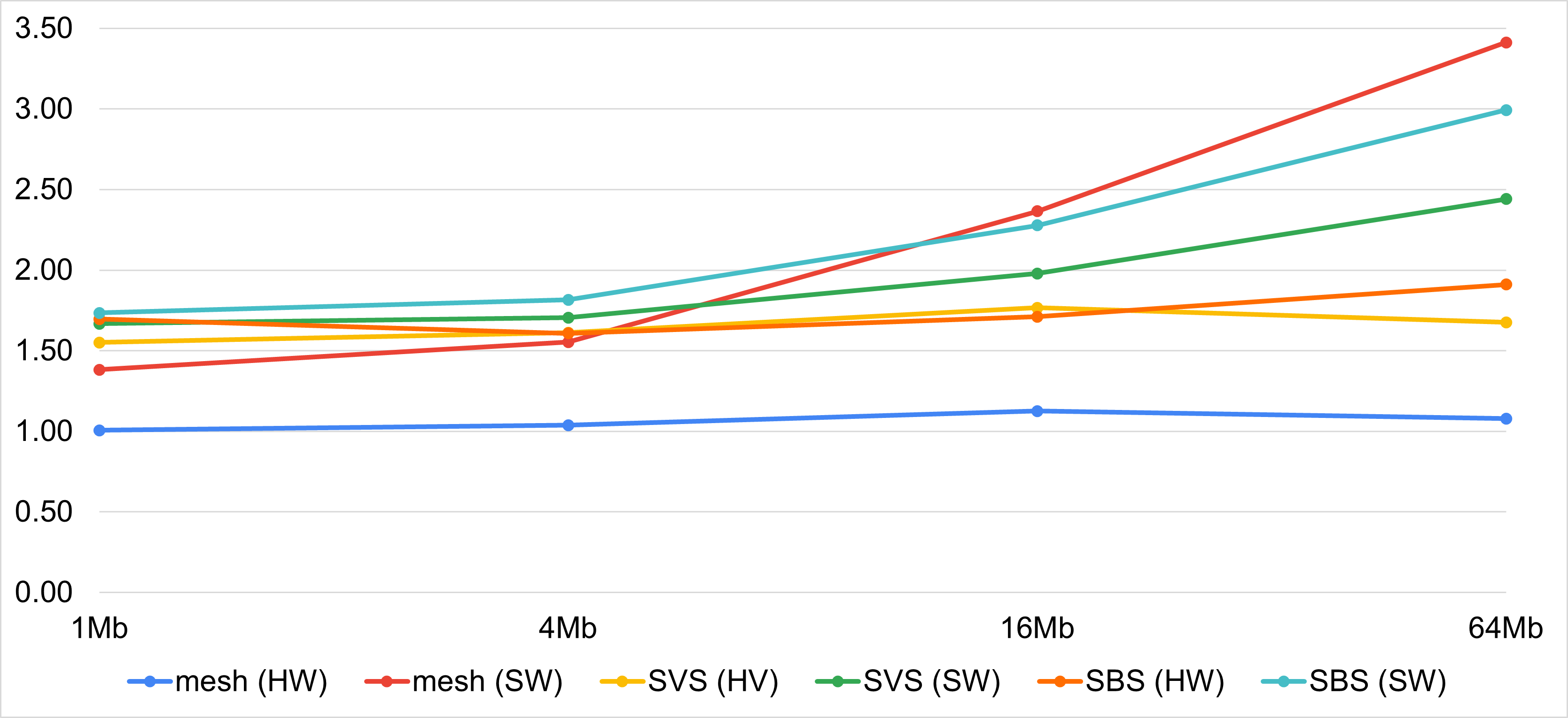}
	\caption{Average rendering time for scenes from figure \ref{fig:sdf_models}. ``SW'' depicts generated in compute shader software implementation. ``HW'' -- generated implementation in the intersection shader via ray tracing pipeline. }\label{fig:sdfperf}
\end{figure}

\begin{figure}[h]
	\includegraphics[width=\linewidth]{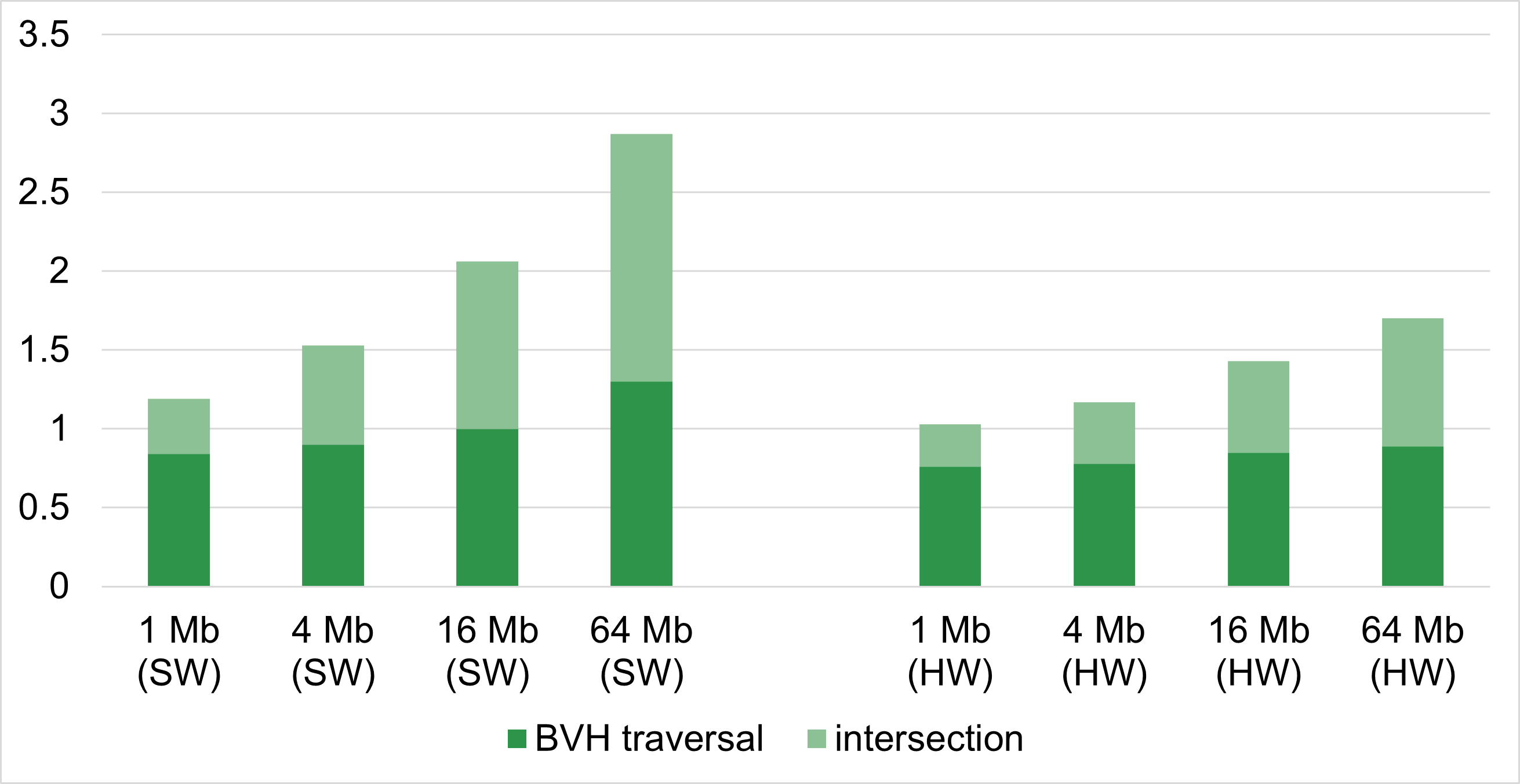}
	\caption{BVH traversal and intersection time for SBS with and without hardware acceleration.}\label{fig:sdfperf2}
\end{figure}

As shown in Figure \ref{fig:sdfperf}, the rendering time for the hardware-accelerated implementation remains nearly constant as the model size increases. This is not the case for the software implementation using a compute shader, where the model size growth increases the tree depth, leading to more divergent ``gather'' memory accesses as described in previous section. Thus, by leveraging hardware acceleration in our programming approach, arbitrary primitives can be added to the polygonal scene without significant performance loss. Moreover, as  Figure \ref{fig:sdfperf2} suggests, hardware acceleration not only speeds up BVH traversal, but also reduces average time for intersection calculataion. Even on GPUs without such acceleration, the rendering algorithm will still work, albeit with slower performance.

%Unlike the work presented in \cite{SDF22}, we aimed not only to understand how quickly distance functions can be rendered using ray tracing but also to evaluate how well they approximate the actual geometry. Therefore, we used triangular meshes as reference geometry and subsequently converted them into SDFs with varying tree depths, and hence, varying memory footprints. This is a significant aspect since the tree depth affects the traversal time. For this reason, we present measurement results for varying tree depths.

%For each 3D model, we rendered 10 viewpoints by rotating the model around its axis at the center of the Cornell Box. The PSNR metric value was averaged across all 10 viewpoints. To enhance the accuracy of time measurement, we performed 100 runs of primary ray tracing for each viewpoint and averaged their results. For path tracing, we measured the time once and averaged it across all 10 viewpoints.

\subsection{Radiance Fields}\label{radiancefields}

We implemented the rendering of ReluFields using BVH tree traversal. All non-empty voxels are fed into the BVH builder, ensuring each voxel is contained within a single tree leaf. Subsequently, we perform BVH tree traversal, sampling exactly once at the center of each voxel, thus slightly improving accuracy compared to the original method (table \ref{table:relupsnr}). 

\begin{table}[h]
	\centering
	\begin{tabular}{|c|c|c|c|c|}
		\hline
		CPU/GPU   & ReLU Fields \cite{ReLUFields} & KiloNeRF \cite{KiloNERF} & Ours & Ours (RTX) \\
		\hline
		i9-14700  & - & - & 249 ms & -  \\
		\hline
		RTX4090   & 1391 ms & 7.9 ms & 8.6 ms & 5.5 ms  \\
		\hline
		AMD 780M  & - & - & 81.0 ms & -  \\
		\hline
		Intel 630 & - & - & 168 ms & -  \\
		\hline
		M3 Pro    & - & - & 44.6 ms & -  \\
		\hline
	\end{tabular}
	\caption{Average rendering time in ms for all scenes (fig. \ref{fig:radiance_fields}) for different GPUs (1024x1024 pixels). The dashes indicate that existing implementations of radiance fields do not work on the specified platforms due to their dependency on CUDA.}
	\label{table:speeup}
\end{table}

Our implementation is nearly 180 times faster than the original CUDA + PyTorch version (table \ref{table:speeup}) due to a combination of algorithmic optimization (our algorithm traverses the BVH tree, while the original uses a constant step) and the capabilities of our programming technology. Specifically, the algorithmic enhancement yields a 17x speedup, and the appropriate backend in our programming technology contributes approximately a 10x speedup.

\begin{table}[h]
	\centering
	\begin{tabular}{|c|c|c|c|}
		\hline
		Scene  & ReLU Fields & Ours & KiloNeRF \\
		\hline
		Ship    & 27.44 & 27.60 & 29.93 \\
		\hline
		Lego    & 29.17 & 29.37 & 35.21 \\
		\hline
		Chair   & 33.82 & 33.93 & 37.26 \\
		\hline
		Hot Dog & 30.12 & 30.30 & 32.82 \\
		\hline
	\end{tabular}
	\caption{PSNR value between reference image and render for different scenes from \cite{ReLUFields}. Thus our implementation of ReLU Fields is correct.}
	\label{table:relupsnr}
\end{table}

\begin{figure}[H]
	\centering
	\includegraphics[width=0.5\linewidth]{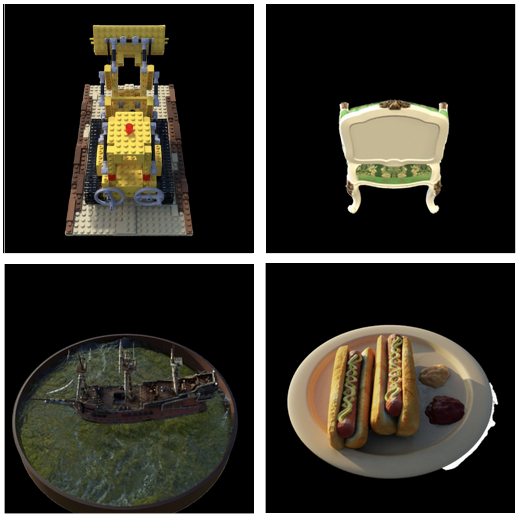}
	\caption{Example of rendered scenes in our tests\label{fig:radiance_fields}}
\end{figure} 

\FloatBarrier

\subsection{Path Tracing with complex material}

\textbf{Performance.} We have developed our own rendering system, Hydra3, which replicates certain material models from Mitsuba3 and PBRT4. Our primary motivation was to create a more lightweight (we have compared complexity in table \ref{tab:locs}), cross-platform, and high-performance rendering system with approximately the same functionality. Therefore, the images of test scenes generated by our system using path tracing are nearly identical to those produced by the PBRT4 and Mitsuba3 rendering systems. We present a table with the PSNR metric between Hydra3 and Mitsuba3 (Table \ref{tab:example_table}), which demonstrates that the differences are insignificant and mainly is a result of Monte Carlo noise (which is significant only for bathroom, bedroom and kitchen scenes). Additionally, we have compared the developed rendering system with LuisaRender, which is based on the Luisa framework. We also have compared Cross$^\dagger$RT to other technologies for kernel generation mode (i.e. plain wavefront vs megakernel). The results are presented in table \ref{tab:rendering_comparison}.

\begin{table}[ht]
	\centering
	\begin{tabular}{|c|c|c|}
		\hline
		Scene & Image resolution & PSNR value\\ \hline
		bathroom & 1280 x 720    & 25 \\ \hline
		bedroom & 1280 x 720     & 31 \\ \hline
		classroom & 1280 x 720   & 38 \\ \hline
		coffee & 800 x 1000      & 39 \\ \hline
		dining room & 1280 x 720 & 36 \\ \hline
		kitchen & 1280 x 720     & 33 \\ \hline
		spaceship & 1280 x 720   & 43 \\ \hline
		staircase & 720 x 1280   & 41 \\ \hline
		staircase2 & 1024x1024   & 36 \\ \hline
	\end{tabular}
	\caption{PSNR values and image resolution for test scenes. PSNR is computed between Hydra3 and Mitsuba3 renders.}
	\label{tab:example_table}
\end{table}

\begin{figure}[H]
	\includegraphics[width=\linewidth]{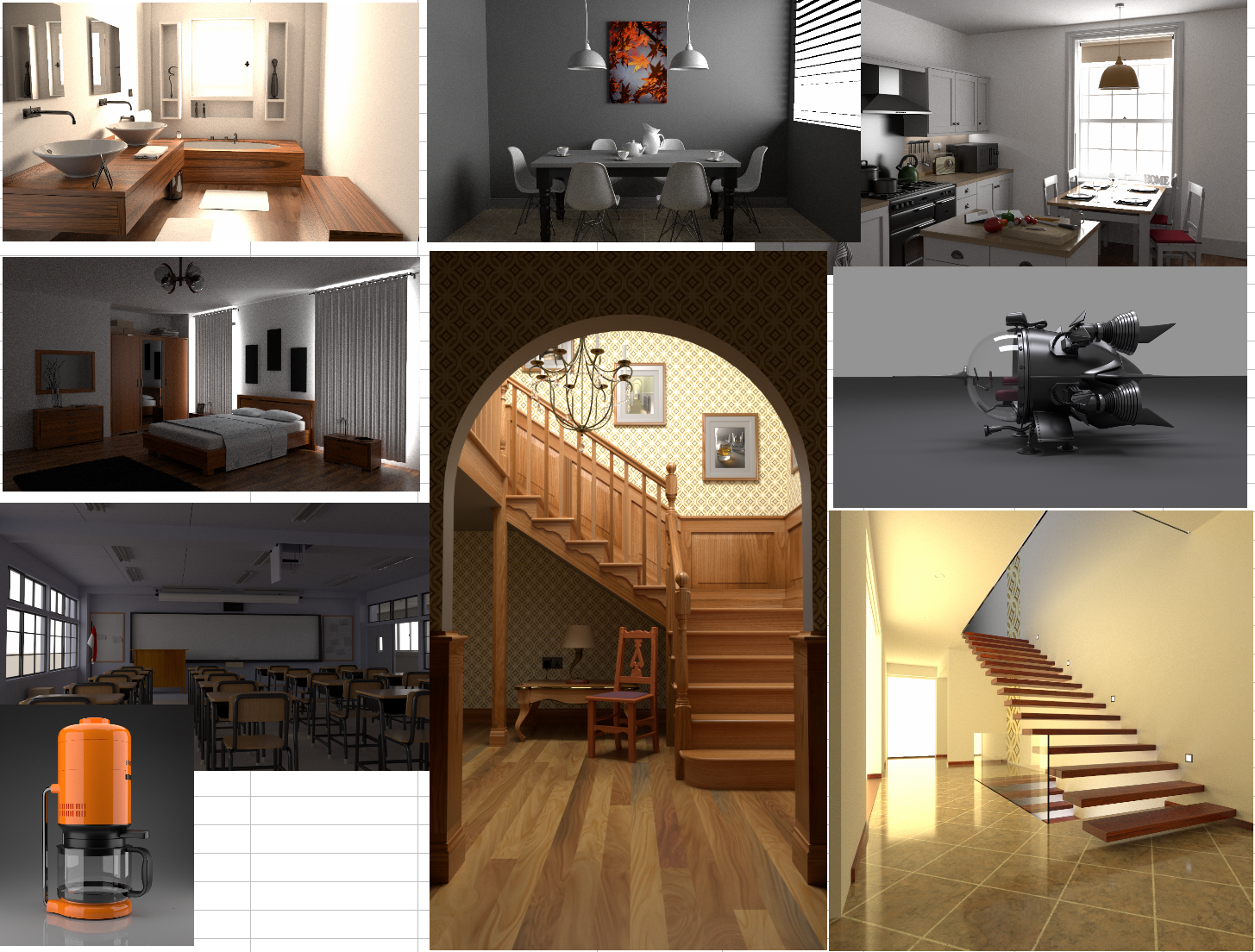}
	\caption{A collage of images from the 3D scenes used for comparison. All scenes were adapted from \cite{scenes}. For comparison with Luisa we used versions of these scenes provided by developers of Luisa on github. \label{fig:allscenes}}
\end{figure}

\begin{table}[ht]
	\centering
	\begin{tabular}{|l|c|c|c|c|c|c|c|}
		\hline
		\textbf{Scene} & \textbf{Ours/M} & \textbf{Ours/W} & \textbf{PBRT4/W} & \textbf{Mi/M} & \textbf{Mi/W} & \textbf{Luisa/M} & \textbf{Luisa/W} \\ \hline
		Bathroom      & \textbf{15.7} & 30.8 & 82.3  & 33.2  & - & 18.1  & 21.8 \\ \hline
		Bedroom     & \textbf{28.6} & 46.2 & 99.2  & 49.1  & - & 44.8  & 50.0 \\ \hline
		Classroom   & \textbf{13.3}  & 32.8 & 77.1  & 44.8  & - & 21.5  & 23.6 \\ \hline
		Coffee      & 6.80  & 14.0 & 60.0  & 15.2  & - & \textbf{6.33}  & 6.70 \\ \hline
		Dining room & \textbf{12.2} & 31.0 & 50.4  & 34.9  & - & 12.8  & 16.9 \\ \hline
		Kitchen     & \textbf{22.2} & 35.4 & 93.1  & 31.2  & - & 24.0  & 25.4 \\ \hline
		Spaceship   & \textbf{8.69}  & 14.2 & 46.8  & 20.8  & - & 9.76  & 10.6 \\ \hline
		Staircase   & 23.8  & 40.5 & 87.9  & 24.1  & - & \textbf{18.0}  & 23.8 \\ \hline
		Staircase2  & \textbf{15.1}  & 29.0 & 122.6 & 33.8  & - & 20.3  & 17.9 \\ \hline
	\end{tabular}
	\caption{Rendering time in seconds across different rendering systems and scenes, 2048 samples per pixel, Nvidia RTX2070 Super. Less rendering time is better. Best result in row is marked with bold text. Columns ``Ours/M'' and ``Ours/W'' are corresponds to Hydra3 via Cross$^\dagger$RT with Megakernel (/M) mode and plain Wavefront (/W) modes. Column ``PBRT4/W'' corresponds to PBRT4 via OptiX and Wavefront approach.  Columns ``Mi/M'' and ``Mi/W'' corresponds to Mitsuba3 via DrJIT and same for Luisa.  }
	\label{tab:rendering_comparison}
\end{table}

\textbf{Complexity.} We have compared different generated variants of virtual function calls in table \ref{tab:virtual_functions}. Our rendering system has less functionality compared to established solutions such as PBRT4 and Mitsuba3 because our initial goal was to develop a lightweight and, as much as possible, straightforward solution. This applies both to the rendering process and to the programming technology. Most of the logic in Hydra3 is implemented within the Integrator class, which is the primary class handled by the proposed CrossRT translator. Users are encouraged to make any substantial modifications to the logic of this integrator in subclasses derived from Integrator. Thus, Hydra3 is not a single large rendering system but rather a collection of smaller, specialized rendering systems tailored to specific tasks. Currently, besides the main class, there are only three hierarchies with virtual functions: the material hierarchy, the light sources hierarchy, and the hierarchy of geometric objects (including volumes) used within the BVHRT class, which inherits from ISceneObject. This is not a limitation of the system architecture or our translator, as additional hierarchies can be incorporated if necessary.

\begin{table}[ht]
	\centering
	\begin{tabular}{|l|c|c|c|}
		\hline
		\textbf{Scene} & \textbf{Compute + switch} & \textbf{RayGen + switch} & \textbf{RayGen + callable}  \\ \hline
		Bathroom    & 15.7 & 33.0 & - \\ \hline
		Bedroom     & 28.6 & 43.1 & - \\ \hline
		Classroom   & 13.3 & 30.7 & - \\ \hline
		Coffee      & 6.80 & 31.2 & - \\ \hline
		Dining room & 12.2 & 31.0 & - \\ \hline
		Kitchen     & 22.2 & 36.2 & - \\ \hline
		Spaceship   & 8.69 & 14.2 & - \\ \hline
		Staircase   & 23.8 & 40.0 & - \\ \hline
		Staircase2  & 15.1 & 31.5 & - \\ \hline
	\end{tabular}
	\caption{Rendering time in seconds across different generated implementations of virtual functions. Megakernel mode, 2048 samples per pixel, Nvidia RTX2070 Super. The column ``Compute + switch'' is responsible for generated compute shader calls where virtual function call is implemented via switch construct like in Luisa. Next column ``RayGen + switch'' is the same as previous except that implementation is generated inside of a so-called "Ray Generation Shader" from ray tracing pipeline. We conducted this measurement to specifically assess the overhead of the ray tracing pipeline. The last column is a hardware-accelerated implementation of virtual functions which is supported only in ray tracing pipeline. }
	\label{tab:virtual_functions}
\end{table}

\begin{table}[ht]
	\centering
	\begin{tabular}{|l|c|c|}
		\hline
		\textbf{Renderer} & \textbf{Lines Of Code} & \textbf{Build time} \\ \hline
		PBRT3        & 440K & 32 s \\ \hline
		PBRT4        & 1.8M & 54 s \\ \hline
		Mitsuba3     & 1.1M & 390 s \\ \hline
		LuisaRender  & 260K & 680 s \\ \hline
		Hydra3 (C++, Core) & 7.5K & 10 s \\ \hline
		Hydra3 (Shaders, generated) & 13K & 0.3s \\ \hline
		Hydra3 (Vulkan, generated)  & 100K & 14.2s \\ \hline
		Hydra3 (everything including libraries) & 136K & 12 s \\ \hline
		Cross$^\dagger$RT (without llvm)        & 15K  & 25 s \\ \hline
	\end{tabular}
	\caption{Complexity estimation of different rendering systems and Cross$^\dagger$RT in lines of code and build time on the Intel Core i9 14900K (14 cores, 28 threads). Hydra3 is our system developed within Cross$^\dagger$RT approach. }
	\label{tab:locs}
\end{table}
\FloatBarrier

\subsection{Expert assessment}

In order to assess the complexity of working with our technology we verified it during a university GPU programming course. Three assignments were issued in the course: (1) tracing an analytical SDF (such as a three-dimensional Mandelbrot fractal), (2) optimizing and rendering a neural SDF from a point cloud and a set of known distances, and (3) ray tracing for radiance fields based on ReLuFields. We did not restrict students in their choice of programming technology; however, for our technology, we provided C++ templates and instructions for using the Corss$^\dagger$RT translator. From the ecosystem, only the LiteMath library for basic mathematical operations was used in the assignments. Thus, a student, having completed the basic part of the assignment on the CPU, could then choose any programming technology (e.g., CUDA) and port the code accordingly. Alternatively, they could run our translator and attempt to compile the generated GLSL shaders and generated code on Vulkan. Statistic is shown in table \ref{tab:coursestat}.

Let us discuss our interpretation of the statistics. The experiment involved graduate students who had previously completed a CUDA course and undergraduate students without GPU experience. Both groups of students completed the first assignment. The second assignment was completed only by the graduate students, while the third assignment was completed only by the undergraduates. As evident, despite many students having prior exposure to CUDA, the repeated experience without a prepared CUDA assignment template proved challenging. Most students, who submitted the second assignment, managed to complete it only on the CPU. Conversely, students without CUDA experience predominantly succeeded in porting the code to the GPU using our programming technology.

Such a comparison is, of course, not entirely fair. However, if we had prepared assignment templates for CUDA, most students would likely have chosen this programming technology for another reason—its widespread use in the industry and hence its higher demand as a skill for employers. Therefore, our comparison at least demonstrates that the effort required to learn Cross$^\dagger$RT is comparable to that of learning CUDA, and under certain conditions (such as prepared templates), it might be even lower. Nearly half of the students who used CUDA reported issues with the installation and configuration of CUDA on Linux. At the same time, two of our students successfully built and utilized our technology on MacOS through MoltenVK.

\begin{table}
\begin{tabular}{|c|c|c|c|c|}
	\hline
	Task                & Cross$^\dagger$RT & CUDA & Other & CPU only \\
	\hline
	№1: Analytic SDF    & 12/38 (30\%) & 9/38 (23\%) & 4/38 (10\%)  & 13/38(34\%)   \\
	\hline
	№2: Neural SDF      & 7/32 (22\%)  & 3/32 (10\%) & 4/32 (12\%)  & 18/32 (56\%)  \\
	\hline
	№3: Radiance Fields & 4/7 (57\%)   & 0/7 (0\%)   & 2/7 (28\%)   & 1/7(15\%)  \\
	\hline
\end{tabular}
\caption{GPU programming course statistics.}\label{tab:coursestat}
\end{table}
\FloatBarrier

%%%%%%%%%%%%%%%%%%%%%%%%%%%%%%%%%%%%%%%%%%
\section{Contribution}\label{contribution}
Our primary contribution lies in enabling hardware acceleration for ray tracing on GPUs where available while maintaining cross-platform compatibility on other GPUs and CPUs. Below, we outline the main differences compared to similar technologies.

\subsection{Difference to previous work}

Unlike OptiX \cite{optix}, our programming technology does not require maintaining separate code for CPU and GPU. Code written for the CPU is transferred to the GPU without the need for additional programs that must be specified in OptiX. Moreover, our generated code runs on GPUs from various vendors.

The key difference from Luisa \cite{Luisa} is the ability to utilize hardware acceleration for ray tracing more broadly (e.g., intersection shader) and the ability for the user to manually substitute unsupported hardware acceleration. Additionally, we generate readable source code for both shaders and C++ to manage Vulkan calls. Therefore, even if some hardware acceleration is not currently supported by our translator, the user can extend it themselves by inheriting from the generated code. In Luisa, as in other programming technologies, this is not permitted.

Our work differs to \cite{UE4Cpp} in several ways: (1) Kerry et al.'s work uses UE4 for the host back-end and is targeted at game developers rather than both general-purpose computing and ray tracing; (2) Kerry et al.'s work does not allow input code to be executed outside their environment on the CPU, which is one of the cornerstone of our approach; (3) Our approach offers a different feature set compared to \cite{UE4Cpp}, especially related to hardware-accelerated ray tracing.

%%%%%%%%%%%%%%%%%%%%%%%%%%%%%%%%%%%%%%%%%%
\section{Conclusions}

In our programming technology, we proposed a solution to the fundamental contradiction between cross-platform compatibility and hardware acceleration in ray tracing applications through the generation and potential user refinement of generated code. On target tasks, we demonstrated that the performance of the generated code is comparable to the performance of code written by experts using similar and widely-adopted programming technologies. Our approach alters the traditional development concept, as users will eventually encounter debugging and modification of the generated code. However, we consider this a norm in modern programming methods and technologies, especially with the advancement of large language model based code generators.

%%%%%%%%%%%%%%%%%%%%%%%%%%%%%%%%%%%%%%%%%%
\begin{adjustwidth}{-\extralength}{0cm}
%\printendnotes[custom] % Un-comment to print a list of endnotes

\reftitle{References}

\end{adjustwidth}
\end{document}